\documentstyle[12pt]{article}
\newcommand{\lwig}{\mbox{\,\raisebox{.3ex}
{$<$}$\!\!\!\!\!$\raisebox{-.9ex}{$\sim$}\,}}
\newcommand{\gwig}{\mbox{\,\raisebox{.3ex}
{$>$}$\!\!\!\!\!$\raisebox{-.9ex}{$\sim$}}\,}
\begin{document}
\begin{titlepage}
\begin{flushright} CERN-TH.6822/93 \\
UCLA 93/TEP/24
\end{flushright}
\renewcommand{\thefootnote}{\fnsymbol{footnote}}
\begin{center}
{\bf \LARGE Cosmic Ray Signatures \\
of Multi-W Processes } \\
\vspace{0.25cm}
\baselineskip=18pt
{{\bf D.A. Morris}\footnote[1]{morris@uclahep.physics.ucla.edu}  \\
University of California, Los Angeles\\
405 Hilgard Ave., Los Angeles, CA 90024,  U.S.A. \\
{\bf A. Ringwald}\footnote[2]{ringwald@cernvm.cern.ch} \\
Theory Division, CERN\\
CH-1211 Geneva 23, Switzerland}
\date{}
\end{center}
\renewcommand{\thefootnote}{\arbic{footnote}}
\setcounter{footnote}{0}
\vspace{.25cm}
\thispagestyle{empty}
\begin{abstract}\normalsize
\noindent
We explore the discovery potential of cosmic ray physics
experiments for Standard Model processes
involving the nonperturba\-ti\-ve
production of $\gwig~{\cal O}(\alpha_W^{-1}) \simeq 30$
weak gauge bosons. We demonstrate an experimental insensitivity to
proton-induced processes and
emphasize the importance of neutrino-induced
processes. The Fly's Eye
currently constrains the largest region of parameter space characterizing
multi-W phenomena
if a cosmic neutrino flux exists at levels suggested
by recent models of active galactic nuclei.
MACRO (DUMAND) can constrain or observe
additional regions by searching for 1--100 (1--10) characteristic
near-vertical (near-horizontal) spatially compact
energetic muon bundles per year.
\vspace{.5cm}
\end{abstract}
\baselineskip=18pt
\begin{flushleft} CERN-TH.6822/93 \\ UCLA 93/TEP/24\\
August 1993
\end{flushleft}
\end{titlepage}

\baselineskip=18pt
\section{Introduction}

\hspace*{\parindent}
Lowest-order perturbative calculations in the Standard Model
for processes involving the production
of $\gwig~{\cal O}(\alpha_W^{-1}) \simeq 30$
weak gauge bosons suggest an explosive (and unitarity violating)
growth of the associated parton-parton cross section above
center-of-mass energies
$ \gwig~ {\cal O}(\alpha_W^{-1} M_W ) \simeq 2.4$~TeV
[1-6]. Not surprisingly, this intriguing circumstance has drawn
considerable interest  in attempts
to either substantiate or dismiss the possibility that the
lowest-order result is a harbinger of spectacularly rich phenomena
which may be observable at the next generation of hadron colliders
(see ref. [7] for an overview). When unitarity
is restored (most likely due to nonperturbative effects) it
is conceivable that the cross section for
multi-W phenomena\footnote{It is understood in the following that by
`multi-W' phenomena we refer to processes producing a total of
$\gwig ~{\cal O}(\alpha_W^{-1}) \simeq 30$ W's {\bf and} Z's.}
is unobservably small. However, despite concentrated theoretical
efforts toward resolving the issue of large versus small cross sections
\cite{MultiPart},
no definitive answer has arisen;
there is a very real possibility that the question may first
be settled experimentally. Given the stakes involved, a quantitative
consideration of experimental constraints on
multi-W production is clearly desirable.

        The high-energy, high-luminosity environment of hadron
colliders such as the proposed Large Hadron Collider (LHC) and the
Superconducting Super Collider (SSC) would be ideal for
observing or constraining multi-W phenomena over a wide range of
energies and cross sections \cite{fa90,ri91b}.
However, until such machines are commissioned,
cosmic rays provide the only access to the required energy
scales. In this paper we adopt a purely phenomenological
approach and explore the feasibility of discovering or constraining
multi-W phenomena in the context of cosmic rays. We consider both
atmospheric and underground phenomena
induced by cosmic ray protons
and neutrinos. Some of the consequences of neutrino-induced phenomena for
future underwater detectors such as DUMAND \cite{dumand}
and NESTOR \cite{nestor} have already been
discussed in refs.~\cite{mo91,be92}.

By adopting a simple parameterization of multi-W phenomena on the
parton level, we evaluate the discovery potential of
various experimental arrangements over a space of possible theories
involving multi-W production. We demonstrate that even if
proton-induced atmospheric multi-W phenomena occur in Nature,
the features of the resulting air showers are unlikely to allow
one to distinguish them from fluctuations
in a much larger background of generic showers.
On the other hand, ultrahigh energy
neutrinos, for which a sizeable flux has recently
been conjectured from sources such
as active galactic nuclei
[15-18],
offer exciting
possibilities for observing or constraining multi-W phenomena.
Subsurface detectors such
as AMANDA \cite{amanda}, DUMAND, MACRO \cite{macro},
NESTOR, and NT-200 \cite{baikal}
can be sensitive to neutrino-induced multi-W phenomena
in some regions of multi-W parameter space.

The structure of the paper is as follows. In sect.~2
we characterize multi-W phenomena by a two-parameter
working hypothesis which frees us from specifying an
underlying (most likely nonperturbative) mechanism for
multi-W production. We also describe the
gross features of multi-W phenomena and present
discovery limits for the LHC and SSC.
We discuss proton-induced and neutrino-induced multi-W phenomena
in sects.~3 and 4, respectively. In each section
we consider a variety of detection techniques and present
discovery limits which may be contrasted with the
superior sensitivity of future hadron colliders. In sect.~5 we
summarize our results and conclude. In an appendix
we outline our quantitative description of multi-W
processes.

\section{Working Hypothesis}

\hspace*{\parindent}
In the absence of a reliable first-principles calculation
of multi-W production, we adopt
a working hypothesis which allows
us to parametrize the essential features
of nonperturbative phenomena without committing ourselves to a specific
underlying mechanism. We model the parton-parton
cross section for multi-W production by
\begin{equation}
\label{eq:workhyp}
\hat \sigma_{\rm multi-W} = \hat \sigma_0\
\Theta
\bigl( \sqrt{\hat s} -\sqrt{\hat s_0 } \bigr) .
\end{equation}
For the purposes of this paper a parton is
any weakly interacting particle (for example, $q$,$\nu$,$e$).
The $\Theta-$function in eq.~\ref{eq:workhyp} embodies an idealized
onset of nonperturbative multi-W production above a
parton-parton center-of-mass threshold energy
of $\sqrt{\hat s_0}$ with a cross section of
$\hat\sigma_0.$ While these definitions are
convenient, it should be kept in mind that the
true behaviour of the parton-parton multi-W cross section
near threshold may be more complicated than  eq.~\ref{eq:workhyp};
in that case $\sqrt{\hat s_0}$ and $\hat\sigma_0$ in eq.~\ref{eq:workhyp}
should be interpreted as effective parameters.

We wish to explore the possibility of observing multi-W phenomena
characterized by parameters in the range
\begin{equation}
\begin{array}{lcccl}
\label{eq:threshold}
\displaystyle{m_W \over \alpha_W } \simeq 2.4~\mbox{TeV} &
\leq &
\sqrt{\hat s_0} &
\leq &
40~\mbox{TeV}, \\
 & & & & \\
\label{eq:partoncro}
\displaystyle{\alpha_W^2 \over m_W^2 } \simeq 100 ~\mbox{pb} &
\leq &
\hat \sigma_0  &
\leq &
\sigma^{pp}_{\rm inel} \times
\left( \displaystyle{1~\mbox{GeV}\over m_W}\right)^2 \simeq
10~\mu\mbox{b}.
\end{array}
\end{equation}
The lower limit of $\sqrt{\hat s_0}$ is suggested by the
energy scale at which perturbation theory becomes unreliable
\cite{co90,gp90,MultiPart} whereas the upper range is of the order of
the sphaleron mass \cite{kl84}. The lower range of
$\hat \sigma_0 $ follows from dimensional arguments, being
characteristic of a geometrical ``weak'' cross section. The upper
range of $\hat \sigma_0$ is a geometrical ``strong''
cross section suggested by analogies between the weak
SU(2) gauge sector and the color SU(3) gauge sector \cite{ri91a}.
Admittedly, our current theoretical understanding of weak interactions
renders $\hat\sigma_0~\gwig ~1~\mu$b an unlikely scenario;
we nevertheless include it in our analysis so that
it may be tested experimentally. For definiteness, we will assume
throughout this paper that $\hat\sigma_0$ refers to the production
of exactly 30 W bosons; allowing for the production of variable numbers
of W's (and Z's and possibly prompt photons)
is straightforward but is an unnecessary complication
at the level of our investigation.

Figure \ref{fi:universal} shows the
multi-W production cross section for
protons and neutrinos (energy $E$) striking stationary
nucleons and electrons. The curves are obtained by
convoluting eq.~\ref{eq:workhyp}
with the corresponding quark distribution functions
inside nucleons (see appendix for details).
The results are universal in the
sense that they have been scaled by
$\hat \sigma_0$ and $\sqrt{ \hat s_0 }$.

The simultaneous production of ${\cal O}(30)$ W~bosons
at future hadron
colliders like LHC or SSC would lead to
spectacular signatures
[9,10].
Since the average number
of charged hadrons (mainly $\pi^\pm$'s) from hadronic W decays is
$\langle n_{\rm ch}^{(W \rightarrow {\rm hadrons} )}
\rangle \simeq 20,$ one
could typically expect
\begin{equation}
30 \times Br(W \rightarrow {\rm hadrons}) \times
\langle
n_{\rm ch}^{(W \rightarrow {\rm hadrons} )} \rangle \simeq 400~\pi^\pm
\end{equation}
in one multi-W event accompanied by $\simeq 400$ photons from the
decay of $\simeq 200$ $\pi^0$'s. The charged hadrons would
have a minimum average transverse momentum of order
\begin{equation}
\label{eq:transverse2}
p_T^\pi \ge {\cal O}
(m_W/30 )\simeq (2-3) \ \mbox{GeV} ,
\end{equation}
if the W bosons are produced without transverse momentum.
Similarly, one could expect $\simeq 5$
prompt muons ($\simeq 3$ from W decays
and $\simeq 2$ from $c$, $b$, or $\tau$ decay)
carrying a minimum average transverse momentum of
\begin{equation}
\label{eq:transverse}
p_T^\mu \ge {\cal O}( m_W/2 )\simeq 40\ \mbox{GeV} .
\end{equation}
A similar situation holds for other prompt leptons such as $e^\pm,
\nu$ etc.
It is hard to imagine that any other process in the Standard Model
can mimic such a final state \cite{ri91b}.

Figure \ref{fi:lhcssc}
shows the regions in $\sqrt{\hat{s}_0} - \hat \sigma _0$
space accessible to the LHC ($\sqrt{s}=14.6~\mbox{TeV};
{\cal L}=10^{34}~\mbox{cm}^{-2}\mbox{s}^{-1}$)
and the SSC ($\sqrt{s}=40~\mbox{TeV};
{\cal L}=10^{33}~\mbox{cm}^{-2}~\mbox{s}^{-1}).$ The contours
correspond to 1 and 10 events (assuming 100\% detection efficiency)
for $10^7$ s of operation. These contours may be used
as a benchmark to evaluate the effectiveness of various cosmic
ray physics experiments for constraining multi-W phenomena.

\section{Proton-Induced Multi-W Processes}

\hspace*{\parindent}
Cosmic ray protons and heavy nuclei constitute a guaranteed
flux of high-energy primaries potentially capable of initiating
multi-W phenomena. In this section we explore the possibility
of exploiting this cosmic flux and isolating multi-W phenomena from
generic hadronic reactions. We restrict our attention to
cosmic ray protons since they provide the dominant flux in terms
of energy per nucleon.

Since the flux of proton-induced multi-W air showers
is anticipated to be small for even the most optimistic
scenarios, one must resort to experiments with large
effective areas and/or long exposure times.
We will discuss three types of relevant experiments:

1) conventional surface arrays which measure the
$e,\gamma,\mu,$ etc., content of air showers
(AGASA
 \cite{akeno},
CYGNUS
 \cite{cygnus},
CASA
 \cite{casa},
EAS-TOP
 \cite{eastop},
EAS-100
\cite{eas100},
HEGRA
 \cite{hegra},
KASCADE
 \cite{kascade}, etc.),

2) underground experiments sensitive to
downward through-going TeV muons
(LVD \cite{lvd},
MACRO \cite{macro},
Soudan-2 \cite{soudan2}, etc.)
or underwater(-ice) experiments
which detect energetic muons by Cherenkov light
(AMANDA \cite{amanda},
 DUMAND \cite{dumand},
 NESTOR \cite{nestor}, NT-200 \cite{baikal}, etc.),
and

3)
the Fly's Eye  \cite{fly}, an optical array
which is sensitive to nitrogen fluorescence light from air
showers.

\subsection{Conventional surface arrays}

\hspace*{\parindent}
Using a network of detector elements, conventional
surface arrays reconstruct the features of an extensive
air shower by interpolating or extrapolating measurements
of a shower's particle content. Among the particles
sampled are photons and electrons with
$E_e,~E_\gamma~ \gwig ~{\cal O}(1$ MeV),
muons with $E_\mu~ \gwig~ {\cal O}(1$ GeV), and, in some experiments,
hadrons with $E_{\rm had}~ \gwig ~1$ GeV.

Consider a surface array (area $A$) sensitive to showers
above a threshold energy $E_{\rm thresh}.$
For proton-induced phenomena it follows from
elementary considerations that
the number of multi-W showers occurring during time $t$ is
\begin{equation}
\mbox{Number~of~multi-W~showers} =
t \, A \, \Omega \,
\int^{\infty}_{E_{\rm thresh}}
\, dE \,\,
\displaystyle{ \sigma^{pN}_{\rm multi-W}(E)
 \over \sigma_{\rm inel}^{pN}(E) }
\, \, j_p(E) \, \,  ,
\end{equation}
where $j_p(E) = dN_p / ( dA dt d\Omega dE )$ is the cosmic proton
flux and $\Omega$ is the solid angle acceptance of the array.
For the cases we consider the inelastic proton-nucleon
cross section $\sigma_{\rm inel}^{pN}$
is dominated by generic QCD interactions and may be approximated by
$\sigma_{\rm inel}^{pN}(E) \simeq {\rm const.} \simeq 100~\mbox{mb}.$
In principle, multi-W production through proton-electron collisions
is also possible. However as illustrated in Fig.~\ref{fi:universal},
the cosmic proton threshold energy
for multi-W production in $p e^-$ collisions is
$m_p/m_e \simeq 1800$ times larger
than the corresponding $p N$ threshold.
Moreover, the $p e^-$ multi-W cross section is at least 100 times smaller
than the $p N$ multi-W cross section; hence we
neglect multi-W production through $p e^-$ collisions.

Figure \ref{fi:pinitiatedarray} shows contours for the number of
proton-induced multi-W air showers at zenith angles
$\theta \le 60^{\rm o}$ striking a 100~km$^2$ conventional
surface array in $10^7~\mbox{s}.$ For our calculations we
use the cosmic proton flux of the Constant Mass Composition (CMC)
model \cite{ConstantMassComposition} (see Fig. \ref{fi:pflux}).
Though the total flux of ultrahigh energy hadronic
cosmic rays is relatively well measured, its composition
({\it i.e.,} percentage of p, Fe, Mg etc.) is somewhat less certain.
Our use of the CMC proton flux introduces an inherent, though not
critical uncertainty in this respect; the CMC flux is a
compromise between proton-rich \cite{ProtonRich} and proton-poor
\cite{ProtonPoor} scenarios. For purposes of illustration we
optimistically assume $E_{\rm thresh} = 1$~PeV which accommodates
all multi-W thresholds above $\sqrt{\hat{s}_0} \ge 2.4$ TeV.
In 100~km$^2$ arrays like AGASA and EAS-100,
inter-detector spacing on the order of .5--1~km makes
$E_{\rm thresh}=100-1000$~PeV more realistic but
does not change our conclusions.

Though the region of the ($\sqrt{\hat{s}_0},\hat{\sigma}_0$) plane
accessible to surface arrays is clearly limited by the absolute
rate of proton-induced multi-W phenomena,
it is instructive to consider how one might distinguish
an air shower of multi-W origin from
a generic hadronic air shower.
For the remainder of this section, we restrict our
attention to the optimistic scenario of
parton-parton multi-W threshold of $\sqrt{\hat{s}_0} = 2.4$~TeV
with $\hat{\sigma}_0 = 10~\mu\mbox{b};$ for this choice of parameters
a 100 km$^2$ array would see approximately 110 (45) multi-W showers
in $10^7$~s at zenith angles $\theta < 60^{\rm o}$ for
$E_{\rm thresh} = 1$ PeV  (100 PeV).
As illustrated in Fig.~\ref{fi:pinitiatedmultiwflux},
the combination of a rising $\sigma_{\rm multi-W}^{pN}(E)$
with a falling cosmic proton flux spectrum
implies that typical multi-W showers would have energies well above the
corresponding cosmic proton threshold energy of
$E_p^{\rm thresh}= \hat s_0/(2 m_p) \simeq 3$~PeV.
The most probable shower energy is $\simeq 30$ PeV and the
average shower energy is $\simeq 250~\mbox{PeV}$ due to a long tail
on the distribution.

Consider the characteristics of the most probable ( $\simeq 30$~PeV)
multi-W air showers.
To phrase our results in experimentally relevant terms
we use the computer program {\sc showersim} \cite{showersim}
to simulate multi-W air showers and generate samples
of generic proton-induced and iron-induced showers
(see appendix for details). Figure \ref{fi:lateraldistsurfacearray}
compares 30~PeV multi-W, proton and iron showers
in terms of radial particle densities
(with respect to a vertical shower axis)
of electrons ($E_e \ge 1~\mbox{MeV}$),
muons $(E_\mu \ge 1~\mbox{GeV}$)
and hadrons ($E_{\rm had} \ge 1~\mbox{GeV}$).
Each curve is averaged over 25--100 showers taking into
account the distribution of the depth of first interaction
in the upper atmosphere. The densities
in Fig.~\ref{fi:lateraldistsurfacearray} correspond to an
an observation depth of $800~\mbox{g/cm}^2$
(roughly the CYGNUS array depth \cite{cygnus}).
The corresponding average particle numbers are listed in Table~1.
We
neglect possible systematic uncertainties
in the experimental determination of shower energies which may be
important in practice.

The differences between the particle density profiles of
30~PeV showers in Fig.~\ref{fi:lateraldistsurfacearray} are
hardly striking. While there are identifiable systematic differences
between average showers of different origin, the differences
do not appear to be sufficient to discriminate between multi-W
showers and fluctuations in generic proton or iron showers.
We emphasize this point by noting that in the CMC flux model,
the differential fluxes of 30~PeV generic proton-induced,
iron-induced and multi-W showers (with
$\sqrt{\hat{s}_0} = 2.4$~TeV, $\hat{\sigma}_0 = 10~\mu\mbox{b}$)
stand in the proportion p : Fe : multi-W
$ \simeq 1.2 \times 10^5 : 1.1 \times 10^5 : 1.$

Considering the spectacular underlying nature of multi-W phenomenon
the similarities between proton-induced multi-W showers
and generic proton and iron may appear surprising. However,
a proton-induced multi-W shower is
actually a superposition of an ``interesting'' prompt shower
component seeded by the instantaneous decays of W bosons
and  a generic ``uninteresting'' component
initiated by the proton fragment which does not participate in
the multi-W production subprocess. The proton fragment emerges from
the region of multi-W production, hadronizes, and subsequently
generates a generic hadronic shower deeper in the atmosphere.
Since the proton fragment typically carries a substantial fraction
of the primary proton energy, the so-called leading particle effect,
a large generic component to a proton-induced multi-W shower
jeopardizes the chances of isolating multi-W showers from
``pure'' generic showers. As shown in Fig. \ref{fi:fraction},
approximately 60\% (20\%) of the total energy in a 30~PeV (5~PeV)
proton-induced multi-W air shower is carried by the generic component.

To minimize the effects of the uninteresting
generic component of multi-W showers and accentuate the prompt
component one may be tempted to consider multi-W showers
close to the relevant threshold energy.
Flatter particle densities near the core of
5~PeV multi-W showers (compared to 30~PeV showers)
in Fig.~\ref{fi:lateraldistsurfacearray} suggest
the larger transverse momentum characteristic of particles from
W decay. However, it is unlikely that one can capitalize
on such effects in practice. Due to the slow turn-on of
$\sigma_{\rm multi-W}^{pN}$ (see Fig. \ref{fi:universal}),
showers just above threshold are rare; for the case at hand
only $.0002\%~(17\%)$ of proton-induced multi-W showers
have energy less than 5 PeV (30 PeV). This corresponds to
to only one proton-induced multi-W air shower
(with $E_{\rm shower} < 5~{\rm PeV}$) in a $100~{\rm km}^2$ array
every 4400 years! In summary, the prospects for
detecting proton-induced multi-W phenomena using conventional
surface arrays are poor.

\subsection{Underground detectors}

\hspace*{\parindent}
In an effort to overcome the complications introduced by a large
generic shower component in proton-induced multi-W air showers,
it is helpful to concentrate on aspects of air showers which
reflect the nature of the primary hard interaction. Energetic
muons produced from the prompt decays of W bosons or from
the weak decays of energetic mesons are good candidates in this
respect. For definiteness, we restrict our attention to muons
with $E_\mu > 1.5~\mbox{TeV};$ such muons can penetrate to deep
underground detectors such as MACRO \cite{macro}.

As shown in Fig. \ref{fi:lateraldisttevmuons} the lateral
distribution of TeV muons in 30 PeV multi-W showers is flatter
than the corresponding distributions for 30 PeV generic p-induced
or Fe-induced showers. The flatter multi-W distribution
is characteristic of the large transverse momentum of the
prompt muons from W (and Z) decays, eq.~\ref{eq:transverse},
and the decays of pions and kaons, eq.~\ref{eq:transverse2}.
The corresponding average number of TeV muons in each type
of shower is given in the last column of Table~1.

Though distinct, TeV muon signatures from proton-induced
multi-W phenomena are limited by small event rates
and practical detector sizes. The event rate contours
for MACRO can be estimated from the 100 km$^2$
surface array contours in Fig. \ref{fi:pinitiatedarray}
simply by scaling by the area ratio
$( 12~ \mbox{m} \times 77~\mbox{m} ) / 100~\mbox{km}^2 \simeq 10^{-5},$
implying negligibly small rates.
Even for poorly motivated parameters such as
($\sqrt{\hat s_0}=~2.4$ TeV, $\hat\sigma_0= 100~\mu$b)
one would need an underground detector with an area
sensitive to downward moving muons of ${\cal O}(10^5$ m$^2$)
to see penetrating muons from one proton-induced
multi-W event in $10^7$ s.
In view of these small rates, present and future underground
detectors are not sensitive to penetrating muon signatures
of proton-induced multi-W phenomena.

\subsection{Fly's Eye}

\hspace*{\parindent}
Finally, we turn to the discovery potential
of the Fly's Eye \cite{fly}, an optical array sensitive to
nitrogen fluorescence light from air showers whose
trajectories do not necessarily intersect the array.
By detecting fluorescence light emitted as air showers
streak across the sky, the
Fly's Eye is capable of reconstructing the longitudinal
development of air showers with energy greater than
$E_{\rm thresh}= 100$~PeV (see
ref. \cite{sok89} for a pedagogical introduction).

Analogous to our previous calculation for conventional arrays, the
number of proton-induced multi-W showers seen in time $t$ by the
Fly's Eye is given by
\begin{equation}
\mbox{Number~of~multi-W~showers} =
t \,
\int^{\infty}_{E_{\rm thresh}}
\, dE \,
\displaystyle{ \sigma^{pN}_{\rm multi-W}(E)
 \over \sigma_{\rm inel}^{pN}(E) }
\, j_p(E) \,
 A \Omega(E) \, \,  ,
\end{equation}
where the acceptance, $A \Omega, $
is a function of energy and hence appears under the integral.
Fig. \ref{fi:pinitiatedarray} shows event number contours
for proton-induced multi-W air showers corresponding to
$10^7$~s operation of the Fly's Eye using the CMC proton flux and the
acceptance of ref.~\cite{fly}. Due to its sensitivity, the Fly's Eye
operates only on clear, moonless nights; approximately
$2\times 10^6$ s of observation time is
possible in one calendar year \cite{fly}.

Despite the limited region of multi-W parameter space
accessible to the Fly's Eye through proton-induced multi-W
air showers we consider whether
multi-W showers, if present, can be differentiated from generic showers.
In the same spirit that one expects generic Fe-induced air showers
to develop more rapidly than generic p-induced showers,
one might expect that the large initial multiplicity
in a multi-W event (from the immediate decay of the W bosons)
leads to an accelerated development of the corresponding air shower.
To test this hypothesis we compare the longitudinal profiles
({\it i.e.}, the number of electrons as a function of shower depth)
of multi-W showers with those of generic p- and Fe-induced air showers.
Fig.~\ref{fi:longdev} shows samples of 150~PeV vertical showers
of each type. The multi-W air showers assume a parton-parton threshold
of $\sqrt{\hat s_0} = 5$ TeV so that 150~PeV showers
are approximately a factor of 10 above
the corresponding proton threshold of
$\hat s_0 / ( 2 m_p) \simeq 13$ PeV
and are not atypical; for this choice of threshold the Fly's Eye
would expect to see 1--10 multi-W air showers within 10$^7$ s observation
time for $\hat \sigma_0 $ in the
range 10-100~$\mu$b.

For ease of comparison in Fig.~\ref{fi:longdev}, the
depth of first interaction of 150~PeV
cosmic ray protons (Fe nuclei)
was fixed at 42~g/cm$^2$ (11 g/cm$^2$) which
corresponds to the average depth of first interaction.
While systematic differences are evident
between profiles of different origin,
the longitudinal profiles of multi-W showers
are not sufficiently distinctive to prevent
confusion with fluctuations in generic air showers.
For ($\sqrt{\hat s_0} = 5$ TeV, $\hat \sigma_0 = 10~\mu$b)
the differential fluxes of 150~PeV generic proton-induced,
iron-induced and multi-W showers stand in proportion to
$\simeq 80000:75000:1$ in the the CMC model.

\section{Neutrino-Induced Multi-W Processes}

\hspace*{\parindent}
A cosmic flux of ultrahigh energy neutrinos would a provide
a novel opportunity to search for multi-W phenomena.
Unlike proton-induced multi-W production which must compete
with ${\cal O}(100$~mb) generic hadronic processes,
neutrino-induced multi-W production competes only
with ${\cal O}$(nb) weak interaction processes.
Fortuitously, considerable enthusiasm has been generated recently
by predictions of a large flux of cosmic neutrinos from
active galactic nuclei (AGN) [15-18].
Indeed, in the model of Stecker {\it et al.}
(see Fig.~\ref{fi:pflux})
the diffuse flux of PeV neutrinos from AGN
is comparable to the flux of cosmic protons in the
CMC model! If only generic charged current interactions are
operative, the Stecker {\it et al.} flux suggests that
DUMAND should see 154(66)
single muon events per year
with energies $E_\mu\geq 100$ GeV (10 TeV)
at zenith angles $\theta > 70^{\rm o} \cite{st92}.$
In this section we explore alternatives for
uncovering neutrino-induced multi-W phenomena
assuming that a flux of ultrahigh energy neutrinos does, in fact, exist.

We restrict our attention to multi-W phenomena in
neutrino-nucleon collisions and neglect multi-W phenomena from
neutrino-electron collisions. As may be deduced from
Fig.~\ref{fi:universal}, the ratio
$\sigma_{\rm multi-W}^{\nu e^-}/\sigma_{\rm multi-W}^{\nu N}$
reaches a maximum of $\simeq 1/30$  when $\sigma_{\rm multi-W}^{\nu e^-}$
turns on. Since the number density of electrons in matter is
nominally half that of nucleons, $\lwig\, 1/60 \simeq 2\%$ of multi-W
events are due to neutrino-electron interactions.

Concerning neutrino flux attenuation due
to competing processes, we neglect
generic $\nu e^-$ weak interactions
compared to generic $\nu N$ weak interactions. In the
energy range of interest the neutrino-electron cross section
due to generic weak interactions is $ \lwig ~{\cal O}(5 \%)$
of the corresponding generic neutrino-nucleon cross sections \cite{be81}
for $\nu_e,\nu_\mu$ and $\bar\nu_\mu.$ The only exception
is the Glashow resonance $\bar\nu_e + e^- \rightarrow W^-$
at $E_{\bar\nu_e} \simeq 6.3$ PeV which is ${\cal O}(10^2-10^3)$
times larger than the generic $\sigma_{\bar\nu_e N}$ \cite{wil85}.
However, since $\bar\nu_e$ are anticipated to make up
only ${\cal O}(1/6)$ of ultrahigh energy neutrinos of AGN
origin and the Glashow resonance is relevant only near
the lowest of multi-W thresholds (before
$\sigma_{\rm multi-W}^{\nu N}$ has fully turned on), we neglect this
effect for multi-W production.

We divide our discussion of neutrino-induced multi-W phenomena
into four sections. Given that AGN are the most plausible
source of our assumed neutrino flux, we investigate in
sect.~\ref{se:agnsection} the
conditions under which large neutrino cross sections for multi-W
production are compatible with AGN neutrino production mechanisms.
In sect.~\ref{se:flyseyeneut}
we discuss constraints on neutrino-induced
multi-W production from the Fly's Eye
and in sect.~\ref{se:conventionalneut}
we consider the possibility of using conventional air
shower arrays. In sect.~\ref{se:subsurfaceneut} we evaluate
the potential of subsurface detectors
to observe contained multi-W phenomena
and discuss the detection of distant multi-W phenomena through
searches for energetic muon bundles.

\subsection{AGN neutrinos with large cross sections}
\label{se:agnsection}

\hspace*{\parindent}
Active galactic nuclei are natural candidates for ultrahigh
energy neutrino production \cite{oldagn,astro}.
In the AGN model of Stecker {\it et al.} \cite{st91} charged
pions are produced in reactions such as $p \gamma \rightarrow
\Delta^+ \rightarrow n \pi^+$
when protons accelerated by a spherical accretion shock \cite{kaz86}
collide with the dense gas of ultraviolet photons
in the innermost region around the central black hole.
Charged pions decay and give rise to neutrinos, whereas
photons produced through $\pi^0$ decay cascade to lower energies,
eventually appearing as X--rays.
If it is assumed that the diffuse X--ray background is primarily from
AGN, the observed X--ray flux can be used to normalize the
calculation of the neutrino flux.
Szabo and Protheroe \cite{sz92} have extended this model
by including pion production through
$p p$ interactions. The AGN model by
Biermann and collaborators \cite{bi92} differs from the
model used in refs. \cite{st91,sz92} mainly in the geometry;
the shocks needed for the acceleration of the protons
are assumed to arise in the bipolar outflow
of gas and plasma perpendicular to the accretion disc \cite{man92}.

Among other considerations, prolific neutrino production by
AGN is a function of the matter density in the vicinity of the
pion production; if the medium is too dense, charged pions
undergo subsequent hadronic interactions instead of decaying to give
neutrinos. For example, for PeV pions the matter density should
be less than $10^{-8}$ g/cm$^3$ \cite{st92}.
Assuming that the prevailing conditions are indeed
conducive to neutrino production, we must consider whether
or not potentially large neutrino-induced multi-W cross sections
permit produced neutrinos to escape. For example, one might worry about
neutrino reabsorption due to multi-W phenomena through
collisions with the dense gas of ultraviolet photons
required by the AGN model of Stecker {\it et al}..
However, even a low parton-parton multi-W threshold of
$\sqrt{\hat s_0} = 2.4$ TeV corresponds to a neutrino
threshold energy (colliding with $\simeq 40$ eV photons)
of $3\times 10^{13}$ GeV; we can safely ignore this effect.
More important is the effect of neutrino reabsorption by matter;
the relevant parameter is the column density seen by particles
escaping from the inner regions of AGN.

In the Stecker {\it et al.} model
the effective escape column density for neutrinos from
the central region of the AGN is of the order of the
column density of the X-ray emitting region,
$X^{\rm Stecker~{\it et~al.}}_{\rm escape} \simeq X_{\rm X-ray}
\simeq {\cal O}(10^{-3} - 10^{-1}$ g/cm$^2$)
\cite{st91,tu89} whereas in the model of
Biermann {\it et al.}
$X^{\rm Biermann~{\it et~al.}}_{\rm escape} \simeq 10^2$ g/cm$^2$
is of the order of a hadronic interaction length \cite{bipr}.
As can be seen in Fig.~\ref{fi:nuabsorption}
which plots the neutrino interaction length
$X_\nu = m_p / \sigma_{\rm multi-W}^{\nu N}(E_\nu),$ significant
reabsorption of neutrinos by AGN is not an issue since
$X_\nu \gg X_{\rm escape}$. A related point, but one which we do not
address in this paper, is the implication
of large neutrino cross sections for stars near the cores of AGN.
Neutrino interaction lengths of ${\cal O}(10^9$ g/cm$^2$)
due to generic charged currents for $E_\nu \simeq 1-100$ PeV
are sufficient to disrupt stellar evolution near the cores of AGN
\cite{st85,st91}; shorter neutrino interaction lengths
implied by multi-W phenomena may have interesting consequences.

For definiteness, we use the (revised)
Stecker {\it et al.} AGN neutrino flux \cite{st91}
in the following
sections for estimates of expected
rates of multi-W phenomena.
It should be noted, however, that
the fluxes calculated in refs. \cite{sz92,bi92}
generally agree\footnote{At lower energies
refs. \cite{sz92,bi92},
which take $pp$ interactions into account,
give considerably larger fluxes
than ref. \cite{st91}
(see, {\it e.g.,} Fig. 11 in ref. \cite{st92})}
with Stecker {\it et al.}
\cite{st91} above .1~PeV, which is the
energy range we are interested in.
In this sense our use of the Stecker {\it et al.} flux
is intended to be representative of
a large class of AGN flux models.

\subsection{The Fly's Eye experiment}
\label{se:flyseyeneut}

\hspace*{\parindent}
Independent of any neutrino flux model,
the Fly's Eye array  \cite{fly}
puts upper limits on the product of the flux times total cross section
for weakly interacting particles
in the range $10^8~\mbox{GeV} \leq E_\nu \leq 10^{11}~\mbox{GeV}$
assuming that such particles initiate
extensive air showers deep in the atmosphere \cite{ba85}.
The limits are deduced from the non-observation of downward-moving
air showers within the Fly's Eye fiducial volume such that the shower
axis is inclined
80$^{\rm o}$ to 90$^{\rm o}$ from the zenith at the point of
impact
on the Earth. Showers meeting these criteria could only have been
initiated by particles typically penetrating more than
3000 g/cm$^2$ of atmosphere before interacting, which excludes
showers initiated by ultrahigh energy photons and hadrons.

Assuming that the weakly interacting particles referred to by the
Fly's Eye are neutrinos, we denote the relevant cross section
by $\sigma_{\rm tot}^{\nu N}(E_\nu)$ which receives contributions from
both multi-W and familiar charged current weak interactions.
The Fly's Eye limits may be summarized by
$(j_\nu \sigma_{\rm tot}^{\nu N})_{\rm Fly's~Eye}$
$\leq 3.74 \times 10^{-42}
\, \times \,
( E_\nu / 1$ GeV)$^{-1.48}$ s$^{-1}$ sr$^{-1}$ GeV$^{-1}$
\cite{mo91,ma90}.
Since these limits neglect the possibility of flux attenuation
in the upper atmosphere due to large inelastic cross sections,
they nominally apply only if
$\sigma_{\rm tot}^{\nu N}(E_\nu) \le 10~\mu$b.
If one considers a particular neutrino flux model
$j_\nu^{\rm model}(E_\nu)$
the Fly's Eye limit excludes regions
in the $(E_\nu,\sigma_{\rm tot}^{\nu N})$ plane bounded by
\begin{equation}
\label{eq:fe}
\begin{array}{lcccl}
\displaystyle{
( j_\nu \sigma_{\rm tot}^{\nu N})_{\rm Fly's~Eye}
               \over
j_\nu^{\rm model}}
 &  < & \sigma_{\rm tot}^{\nu N}(E_\nu) & < & 10~\mu{\mbox{b}}\, , \\
  & & & & \\
10^8~{\mbox{GeV}} &  < &  E_\nu & < &   10^{11}~{\mbox{GeV}}  .
\end{array}
\end{equation}
If we use the AGN neutrino flux
of Stecker {\it et al.} \cite{st91}
(i.e., set $j_\nu^{\rm model} = j_{\nu}^{{\rm Stecker}~et~al.}),$
the Fly's Eye excludes the hatched region of
Fig.~\ref{fi:neutrinoinitiatedflyseye1}.

In order that $\sigma_{\rm tot}^{\nu N}(E_\nu)$ avoid
the region excluded by eq.~\ref{eq:fe},
only certain combinations of $\sqrt{\hat s_0}$
and $\hat\sigma_0$ are consistent. For example, as
shown in Fig.~\ref{fi:neutrinoinitiatedflyseye1},
for $\sqrt{\hat s_0} = 8$ TeV the range
$.5~\mu\mbox{b} < \hat \sigma_0 < 81~\mu\mbox{b}$ is excluded.
Similar considerations for other values of $\sqrt{\hat s_0}$ result
in the excluded region labeled ``AGN $\nu$'' in
Figure \ref{fi:neutrinoinitiatedflyseyeexcluded}. As may be
quickly verified, the upper left
boundary of the excluded region in Figure
\ref{fi:neutrinoinitiatedflyseyeexcluded} corresponds to limiting
situations in which $\sigma_{\rm tot}^{\nu N}(E_\nu = 10^8$ GeV)
$= 10~\mu$b. In principle the $10~\mu$b
upper bound on the neutrino-nucleon cross section
in eq. \ref{eq:fe} could be enlarged by taking
into account flux attenuation in the upper atmosphere,
which has been neglected in ref. \cite{ba85}. As a consequence, one
could most likely extend the excluded region in
Fig.~\ref{fi:neutrinoinitiatedflyseyeexcluded} into the upper left hand
corner which, taken literally, is not constrained by the Fly's Eye.
A further improvement of the Fly's Eye limit
$(j_\nu \sigma_{\rm tot}^{\nu N})_{\rm Fly's~Eye}$,
by a factor of 10--50,
is expected from the High Resolution (HiRes) Fly's Eye~\cite{fly2}.

Active galactic nuclei are not the only conjectured
sources of ultrahigh energy neutrinos.
As shown in Fig.~\ref{fi:pflux}, when the proposed
AGN neutrino flux dies off beyond $\simeq 1$ EeV,
the dominant component to the neutrino flux
may be due to protons scattering inelastically
off the 2.7~K cosmic background radiation (CBR) \cite{gr66},
producing charged pions that subsequently decay and produce neutrinos
\cite{be76,hi83}. The photoproduced neutrino flux,
$j_\nu^{\rm 2.7~K},$ shown in Fig.~\ref{fi:pflux}
is taken from ref.~\cite{st91}. It is amusing to consider how
the Fly's Eye constraints on neutrino-induced multi-W
production are modified if we account for the possibility of such
photoproduced neutrinos.
If one takes $j_\nu^{\rm model} = j_\nu^{{\rm Stecker}~et~al.} +
j_\nu^{\rm 2.7~K}$ in eq.~\ref{eq:fe}, the Fly's Eye excludes the
$(E_\nu,\sigma_{\rm tot}^{\nu N})$ region shown
Fig.~\ref{fi:neutrinoinitiatedflyseye2} and enlarges the
excluded region in $(\sqrt{\hat s_0},\hat\sigma_0)$ space
by the area labelled ``2.7~K Photoproduced $\nu$'' in
Fig.~\ref{fi:neutrinoinitiatedflyseyeexcluded}.

Though the appearance of an enlarged excluded region
is welcome, it is sensitive to details of the assumed CBR flux.
Had we assumed a CBR flux component $j_\nu^{\rm 2.7~K}$
which was a factor of ten smaller than that shown in Fig.~\ref{fi:pflux}
(corresponding to a lower redshift), the quantity
$(j_\nu \sigma_{\rm tot}^{\nu N})_{\rm Fly's~Eye}
/ j_\nu^{\rm model} $ (corresponding to the solid curve
in Fig.~\ref{fi:neutrinoinitiatedflyseye2}) would not have
dipped below $10~\mu$b and thus would not have
introduced a constraint. We should keep such
uncertainties in mind to avoid attaching undue
significance to the excluded regions in
Fig.~\ref{fi:neutrinoinitiatedflyseyeexcluded}.
Nevertheless it is intriguing to speculate about
detecting CBR neutrinos via multi-W processes since
the prospects for detecting such neutrinos through generic
weak interactions is poor unless the CBR neutrino
flux is associated with a very large redshift.

\subsection{Conventional air shower arrays}
\label{se:conventionalneut}

\hspace*{\parindent}

Consider a conventional air shower array
(area $A$) which is located at
an atmospheric depth $X_0$ and is sensitive to
showers above a threshold $E_{\rm thresh}.$
For showers not close to the horizon it is
straightforward to show that the number of
multi-W showers in time $t$ occurring
in the atmosphere above the detector is
\begin{equation}
\label{eq:neutrinoinducedsurfacearray}
\mbox{Number~of~multi-W~showers} =
{t \, A\over m_p}  \,
\int^{\infty}_{E_{\rm thresh}}
\, dE \, d\Omega \, \,
\displaystyle{ X_0 \over \cos \theta } \,
\sigma^{\nu N}_{\rm multi-W}(E)
\, j_\nu(E) .
\end{equation}
The solid contours of Figure~\ref{fi:neutrinoinitiatedsurfacearray}
correspond to neutrino-induced multi-W events in $10^7$~s
above a 100~km$^2$ array which is
sensitive to showers above 100~PeV within 60$^{\rm o}$ of the
zenith; the AGN neutrino flux of Stecker {\it et al.} \cite{st91}
has been assumed. Even though the neutrino flux is smaller
than the cosmic proton flux in the
CMC model (see Fig. \ref{fi:pflux})
the contours for neutrino-induced
multi-W phenomena cover a considerably larger
region in the $\sqrt{\hat s_0}- \hat\sigma_0$ plane
than the corresponding contours for
proton-induced air showers
(see Fig.~\ref{fi:pinitiatedarray}).
This is due to a more rapid growth of the multi-W cross section
for $\nu N$ scattering
compared to $p N$ scattering (Fig. \ref{fi:universal})
and also due to a much smaller
competing cross section from generic charged current interactions,
$\sigma^{\nu N}_{\rm c.c.} \simeq {\cal O}(1$ nb).

The solid contours in Fig. \ref{fi:neutrinoinitiatedsurfacearray}
do not account for the efficiency of an array to trigger
on low altitude air showers. Such considerations are
crucial for neutrino-induced phenomena since the
distribution of neutrino interactions essentially follows
the density profile of the atmosphere. In an exponential
atmosphere neutrino-induced air showers may be initiated so close
to the array that the showers do not spread out sufficiently
to trigger the array.
Rather than confine ourselves to a detailed analysis of triggering
requirements, consider the following approximation. Suppose
that an array does not trigger on showers initiated
with $500$ g/cm$^2$ of the detection level.
This assumption is reasonable for
vertical showers but is somewhat pessimistic for showers
at larger zenith angles. Contours for ``triggerable''
neutrino-induced multi-W air showers follow from
eq.~\ref{eq:neutrinoinducedsurfacearray} if, in the
integrand, we replace $X_0/\cos\theta$ with
$(X_0 / \cos\theta - 500$ g/cm$^2$).

The dashed contours
of Fig.~\ref{fi:neutrinoinitiatedsurfacearray} correspond to
``triggerable'' neutrino-induced multi-W showers. It is interesting
to note that the contours for $\simeq 5-10$ events
in $10^7$ s for a 100 km$^2$
surface array roughly coincide with the
lower boundary of the Fly's Eye excluded region labelled ``AGN $\nu$'' in
Fig.~\ref{fi:neutrinoinitiatedflyseyeexcluded}; this effect is easily
understood in terms of the relevant acceptances and
exposure times. If, in addition to the AGN neutrino flux of
Stecker {\it et al.}, we were to assume contributions from
2.7~K photoproduced neutrinos as in the previous section,
we would find contours for ${\cal O}(1)$ event in $10^7$ s
for a 100~km$^2$ surface array
which roughly coincide with the lower boundary of the
Fly's Eye excluded region labelled ``2.7 K Photoproduced $\nu$'' in
Fig.~\ref{fi:neutrinoinitiatedflyseyeexcluded}. In other words, assuming
the same neutrino flux, the sensitivity of the Fly's Eye to
multi-W phenomena is comparable to that of a 100 km$^2$ surface
array. For this reason we will not discuss the characteristics of
neutrino-induced multi-W air showers relevant to surface arrays.
Mrenna \cite{mr92} has compared the the features of neutrino-induced
air showers and generic air showers in the context of composite
models \cite{do87} where hypothesized colored
subconstituents of PeV neutrinos interact with
typical QCD cross sections.

Conventional air shower arrays can also search for showers close
to the horizon. Data from the AKENO array places limits
on the existence of electromagnetic (muon-poor) horizontal air
showers initiated deep in the atmosphere \cite{ak71,Zas724}.
Since the AKENO data applies only to shower energies lower than
the multi-PeV range in which we are interested \cite{vasq},
we have not investigated its implications for multi-W phenomena.
Exploitation of the horizontal shower limits would require
consideration of distant neutrino-induced multi-W processes in which
only prompt muons penetrate the intervening atmosphere and
initiate electromagnetic cascades close the surface array.

\subsection{Subsurface experiments}
\label{se:subsurfaceneut}

\hspace*{\parindent}
        Detectors deep below the surface of the Earth, be they
shielded by rock (LVD \cite{lvd}, MACRO \cite{macro},
Soudan-2 \cite{soudan2} etc.), water (DUMAND \cite{dumand},
NES\-TOR \cite{nestor}, NT-200 \cite{baikal}) or ice (AMANDA \cite{amanda})
offer a unique perspective on neutrino-induced phenomena. In this
section we investigate two possible modes for detecting
neutrino-induced multi-W phenomena using subsurface experiments.
We first consider the prospects for
observing contained neutrino-induced multi-W phenomena
and later turn to the
detection of muon bundles arising from neutrino interactions
in the surrounding medium.

        Aside from the energy involved, contained
neutrino-induced multi-W production would reveal its
origins by its enormous multiplicity
(${\cal O}$(400) charged hadrons, ${\cal O}(400)$ photons,
and a few prompt muons and electrons).
Generic deep inelastic $\nu N$ scattering and the
resonant process $\bar\nu_e + e^- \rightarrow W^-
\rightarrow {\rm hadrons}$ can also give contained hadron
production, but only with significantly lower multiplicity.

The number of neutrino-induced multi-W events occurring inside
a subsurface detector volume $V$ during a time $t$ is
\begin{eqnarray}
\lefteqn{\mbox{Number of multi-W events} =}
\qquad \qquad \qquad \nonumber \\
& &
t \,  { \rho \, V\over m_p}
\, \sigma^{\nu N}_{\rm multi-W}
\int  \, dE_\nu \, d\Omega \, j_\nu(E_\nu)
\ \mbox{e}^{- \sigma_{\rm tot}^{\nu N} X(\theta,\phi)/m_p} ,
\end{eqnarray}
where $X(\theta,\phi)$ is the column density of material in
the $(\theta,\phi)$ direction between the detector
and the upper atmosphere and $\rho$ is the density
of the material in which the neutrino interaction occurs.

Figure \ref{fi:neutrinoinitiatedcontained}
shows contours for contained multi-W events in $10^7$~s
in a 1~km$^3$ volume of water at an ocean depth of
4.5~km. This arrangement approximates the
proposed SADCO acoustic array \cite{sadco}
which, though designed to use acoustic techniques to detect the
resonant process $\bar\nu_e + e^- \rightarrow W^-,$
would also be sensitive to multi-W phenomena
which are more energetic.
Acoustic techniques have also been considered for
AMANDA \cite{pr93} and DUMAND \cite{le79}.
The contours in Fig.~\ref{fi:neutrinoinitiatedcontained}
consider the AGN neutrino flux of
Stecker {\it et al.} \cite{st91}
as well as the sum of the Stecker flux with the 2.7~K
photoproduced component from Fig.~\ref{fi:pflux}.

Contours for the number of contained multi-W events in DUMAND can
be obtained from Fig.~\ref{fi:neutrinoinitiatedcontained}
simply by scaling the appropriate volume ratio. If we idealize DUMAND
as a 100~m $\times$ 100~m $\times$ 250~m volume under 4.5~km
of water (neglecting the possibility that its effective acoustic
volume can be larger than its geometrical size), the volume ratio
is $\simeq 1/400.$ Due to a numerical coincidence, the
corresponding contours for MACRO (which we idealize as a
77~m $\times$ 12~m $\times$ 9~m volume at a depth of 3700~hg/cm$^2$
below the surface of a spherical Earth
with $\rho = 2.6$ g/cm$^3$) may be obtained by from
Fig.~\ref{fi:neutrinoinitiatedcontained}
by scaling by a factor of $\simeq 1/40000,$
implying MACRO's insensitivity to contained events.

Due to the enormous
energies involved, one need not to  concentrate on completely
contained multi-W reactions. Of particular interest is the ability for
subsurface detectors to detect muons which arise from
energetic neutrino interactions up to a few kilometers
away. For distant multi-W production the effects of producing
hundreds of hadrons and photons will have died off well before
reaching the detector but the anticipated 2--3 muons from prompt
W decays produced with $E_{\mu} \simeq {\cal O}(100$ TeV)
and $p_T^\mu \simeq {\cal O}(40$ GeV) propagate great distances.
The signature of multi-W production in this case would be
energetic muon bundles.

The ability to detect muons from distant neutrino reactions
increases a subsurface detector's effective neutrino target
volume dramatically and is the premise upon which such
detectors can act as neutrino telescopes.
Considerable effort has recently been directed towards
the prospects of detecting ultrahigh energy neutrinos (most
likely from AGN) using subsurface detectors \cite{st92}.
Despite their limited sensitivity to such phenomena,
Fr\'ejus \cite{me92} and Soudan--2 \cite{al92} have
already placed useful observational constraints
on AGN flux models.

As discussed in ref.~\cite{mo91},
near-horizontal muon bundles in DUMAND and MACRO would
be characteristic of neutrino-induced multi-W phenomena. By concentrating
on large zenith angles, one can avoid the complications from
a large background of muon bundles from generic hadronic
interactions in the atmosphere.
The number of muon bundles containing $k$ muons
detected during time $t$ by a subsurface detector
of length $L$, width $W$ and height $H$ is
\begin{eqnarray}
\label{eq:muonbundlerate}
\lefteqn{ \mbox{Number of muon bundles} = }
\qquad\quad
\nonumber\\
& & t \times \int d\cos\theta \,
{dN_{k\mu}\over dA\ dt\ d\Omega}\, \left[
\displaystyle{2\over \pi } H (L+W) \sin\theta + L W |\cos\theta| \right] ,
\end{eqnarray}
where the quantity in square brackets is the azimuthally
averaged projected area of the detector. The calculation of the
differential flux of muon bundles containing $k$ muons,
$dN_{k\mu} / ( dA\ dt\ d\Omega )$, employs the techniques of
ref.~\cite{mo91} which are summarized in the appendix.

We present in Fig.~\ref{fi:neutrinoinitiatedmacro} contours
for muon bundles  beyond zenith angles of $80^{\rm o}$ for
MACRO and DUMAND for the AGN neutrino flux of Stecker
{\it et al.} \cite{st91}.
Due to our assumed production of 30 W bosons, each muon bundle
consists of approximately 3 muons. The average muon energy
$\langle E_\mu \rangle$
entering the detector and
the average inter-muon separation $\langle r_\mu\rangle$
depend on $\sqrt{\hat s_0}$ and $\hat\sigma_0.$
For example, for $(\sqrt{\hat s_0} = 4$ TeV,
$\hat\sigma_0=10$ nb)
one expects $\simeq 1.5$ bundles per $10^7$~s in DUMAND
with $\langle E_\mu \rangle \simeq {\cal O}(180$ TeV)
and $\langle r_\mu \rangle = {\cal O}(2.5$ m);
for $(\sqrt{\hat s_0} = 4$ TeV, $\hat\sigma_0=1~\mu$b) one
expects $\simeq 30$ bundles per $10^7$~s in DUMAND
with $\langle E_\mu \rangle \simeq {\cal O}(70$ TeV)
and $\langle r_\mu \rangle = {\cal O}(3.6$ m).
Assuming an additional 2.7~K photoproduced neutrino flux
component at the level shown in Fig.~\ref{fi:pflux}
changes the contours of Fig.~\ref{fi:neutrinoinitiatedmacro}
by a negligible amount.

It may also be possible to constrain multi-W phenomena
by searching for non-horizontal muon bundles and thereby enlarge
the accessible region in $\sqrt{\hat s_0}$--$\hat \sigma_0$ space.
Fig. \ref{fi:dumandallzenith}
(Fig. \ref{fi:neutrinoinitiatedallzenith}) shows contours for
muon bundles for zenith angles between 0$^{\rm o}$ and 180$^{\rm o}$ for
DUMAND (MACRO\footnote{Preliminary rates for muon bundles in MACRO
presented in ref.~\cite{mo93} included only AGN $\nu_\mu-$induced multi-W
processes and hence are smaller than those of
fig.~\ref{fi:neutrinoinitiatedallzenith} by a factor of 3.})
for the Stecker {\it et al.} AGN neutrino flux.
An additional 2.7~K photoproduced neutrino flux component
at the level of fig.~\ref{fi:pflux} changes
the 1-10 event contours for DUMAND but has a negligible effect on the
MACRO contours. Since DUMAND is specifically not optimized for downward
muons the DUMAND rates in Fig.~\ref{fi:dumandallzenith}
are presented as a matter of completeness rather than practicality.
MACRO, however, is sensitive to downward muons. Whereas
the inter-muon separation
expected from generic hadronic interactions high in the atmosphere
is typically of ${\cal O}(5$--10 m), neutrino-induced multi-W phenomena
occur primarily inside the Earth and  result in much more spatially
compact muon bundles.

Figure \ref{fi:macrodata} compares MACRO data for
pair-wise muon separation to the contribution expected from
neutrino-induced multi-W phenomena for $(\sqrt{\hat s_0}=2.4$ TeV,
$\hat\sigma_0 = 10~\mu$b). The MACRO
data is taken from Fig.~4 of ref.~\cite{bun92} and corresponds to
muon bundles at zenith angles $\theta < 60^{\rm o}$
detected by two supermodules operating for 2334.3 hours.
The MACRO data contains contributions from muon bundles of
all multiplicities; approximately half of the reconstructed
pairs come from $n_\mu =2$ muon bundles. We suggest that by
separately examining the pair-wise muon separation in
bundles with fixed numbers of muons ({\it e.g.,} $n_\mu$=3)
as has been done the Fr\'ejus collaboration \cite{frejus},
MACRO may be able to put constraints on the existence of
multi-W phenomena. A particularly useful signature of multi-W processes
in this respect is the energy carried by each muon. Muons
arising from multi-W processes in Fig.~\ref{fi:macrodata} would
have energies of approximately 80~{\rm TeV} as they enter the detector
and may be distinguished by mechanisms such as catastrophic energy
loss \cite{me92,al92}. Though some of the region
in $(\sqrt{\hat{s}_0},\hat\sigma_0)$ space to which MACRO is sensitive
is already excluded by the Fly's Eye (assuming the same AGN neutrino
flux), valuable independent constraints may already be possible from
existing MACRO data.

\section{Summary and Conclusions}

\hspace*{\parindent}
Future hadron colliders such as the proposed SSC or LHC offer the
best prospects for observing or constraining multi-W phenomena.
A naive measure of an experiment's sensitivity to multi-W phenomena
is the size of the region in $(\sqrt{\hat s_0},\hat\sigma_0)$
parameter space accessible to the experiment by requiring
at least one multi-W event in $10^7$~s of operation.
By this standard the SSC covers the most territory
(see Fig.~\ref{fi:lhcssc}). For example, at the SSC
one would expect ${\cal O}(100$ events / 10$^7$ s)
if multi-W processes were characterized by
$ ( \sqrt{\hat s_0}=~20$ TeV, $\hat\sigma_0 = 1$ pb).
Even in the relatively noisy environment of a high-luminosity
hadron collider, the spectacular signature of $\gwig~{\cal O}(30)$
gauge bosons in a single event has
no conceivable Standard Model background.
In this sense, when applied to a hadron collider, even a
naive measure of the sensitivity to multi-W phenomena is appropriate.

Before the commissioning of hadron supercolliders,
cosmic ray physics suggests alternative techniques
for searching for multi-W processes induced either
by protons or neutrinos. Taken at face value, the most optimistic
cosmic ray constraints on multi-W phenomena come from
1) the non-observation of neutrino-induced air showers
by Fly's Eye (see Fig.~12) which covers the full range of
$\sqrt{\hat s_0}$ up to 40~{\rm TeV} if
$\hat\sigma_0~\gwig~{\cal O}(10-100$ nb) and
2) searches for horizontal muon bundles in DUMAND (see Fig.~16)
whose sensitivity extends to $\hat\sigma_0
\simeq {\cal O}(1$ nb) if
$\sqrt{\hat s_0}~\lwig~{\cal O}(4$ TeV).
Additional constraints on neutrino-induced phenomena
may be forthcoming from limits on the occurrence of energetic,
spatially compact non-horizontal muon bundles in MACRO
or, more speculatively, from large energy deposits in
proposed acoustic arrays.

Understandably, none of the neutrino-based constraints presented
in this paper are conclusive: all presume the existence of a sizeable
flux of ultrahigh energy neutrinos from AGN or neutrinos photoproduced
off the 2.7~K cosmic background radiation. If it should happen
the required neutrino flux is absent, no conclusions
may be drawn regarding the existence of multi-W processes --- one
would then have to wait for the advent of supercolliders to observe
or exclude multi-W phenomena. However, it is instructive to consider an
intermediate scenario in which a flux of ultrahigh energy neutrinos
is detected in the future but is found to have interactions consistent
generic charged current processes; this too may place constraints
on the existence of exotic phenomena. In any case, since no experiment
has yet studied the interactions of neutrinos with energies greater
than a few hundred GeV, it may be premature to dismiss the
possibility that PeV neutrinos have novel interactions.

To avoid the additional uncertainty of whether an ultrahigh energy
neutrino flux exists, we have also considered the possibility of multi-W
phenomena induced by a cosmic protons. Unfortunately,
the tradeoff for a relatively reliable
proton flux is an overwhelming competing cross section due to
generic hadronic processes. Large competing cross sections
complicate matters in two ways. First, they drastically
reduce the proton flux effectively available for multi-W phenomena.
Even by our naive measure of sensitivity to multi-W phenomena,
a 100 km$^2$ surface array operating for $10^7$ s could only see
multi-W phenomena if $\sqrt{\hat s_0} \lwig {\cal O}(10-12$ TeV)
and $\hat\sigma_0  \gwig {\cal O}(.1-1~\mu$b)
(see, {\it e.g.}, Fig.~\ref{fi:pinitiatedarray}). Second,
large competing cross sections are the source of an
overwhelming background of generic air showers.
Given that there are optimistically
${\cal O}(10^4-10^5)$ generic showers
for every shower of multi-W
origin, we have emphasized that multi-W showers
could easily be mistaken for background fluctuations.
Taking this difficulty into consideration, the true
sensitivity of conventional surface arrays
to proton-induced multi-W phenomena is negligible.
Similar conclusions apply to other techniques
for proton-induced multi-W phenomena such as detecting
downward moving underground TeV muons in MACRO,
or searching for the anomalous longitudinal
development of air showers with the Fly's Eye.

The short term outlook for constraining or detecting
multi-W phenomena in cosmic ray physics is mixed.
Without making additional assumptions (such as
assuming the existence of a large cosmic neutrino flux)
one must focus on proton-induced processes and
conclude that current and future experiments are
effectively insensitive to multi-W phenomena
over the entire range of parameter space where they
might plausibly exist. From this viewpoint one must
wait for terrestrial supercolliders before conclusive
constraints on multi-W processes are established.
While this conservative scenario may very well be true,
an exciting alternative exists. If a sizeable flux of
cosmic neutrinos is present, not only may AMANDA,
DUMAND, MACRO, NESTOR and NT-200 be sensitive to them through
generic weak interactions, but such detectors may also indicate
whether multi-W processes are real or an artifact of our
imperfect understanding of multi-TeV weak interactions.

\section{Acknowledgements}
\hspace*{\parindent}
We would like to thank the following individuals for their
advice and assistance during the course of this work: P. Biermann,
S. Billers,
C. Bloise,
J. Cobb,
M. Goodman,
G. Heinzelmann,
P. Litchfield,
H. Meyer,
S. Mrenna,
A. Shoup,
F. Stecker,
and R. V\'azquez.
We appreciate the aide of the UC Irvine Physics Group.
D.A.M. acknowledges the hospitality of the CERN
theory group during a productive visit. D.A.M. is supported by
the Eloisatron project.

\newpage
\section{Appendix}
\hspace*{\parindent}
In this appendix we outline our quantitative description
of multi-W processes. In sect.~\ref{se:proton} we state
the assumptions and approximations used to model proton-induced
multi-W production and in sect.~\ref{se:showersim} we briefly
describe our simulation of extensive air showers.
Finally, in sect.~\ref{se:muon} we review the calculation
of subsurface detection rates for muon bundles from
neutrino-induced multi-W processes.

\subsection{Proton-induced multi-W processes}
\label{se:proton}
\hspace*{\parindent}
Within our working hypothesis the
proton-nucleon multi-W cross section is given by
\begin{eqnarray}
\label{eq:cross}
\lefteqn{ \sigma_{\rm multi-W}^{pN} =
} \nonumber \\
& & \sum_{ij} \int dx_1 \, dx_2 \,
{\displaystyle
f_i(x_1) \, f_j(x_2)~
+      f_i(x_2) \, f_j(x_1)~
\over
1 + \delta_{ij} }
\hat{\sigma}_0 ~\Theta
\left( \sqrt{x_1 x_2 s} - \sqrt{ \hat{s}_0 } \right),
\nonumber \\
& &
\end{eqnarray}
where $f_i(x)$ is the parton distribution function corresponding to
a parton of flavour $i$ carrying a proton momentum fraction $x.$
The sum extends over all distinct\footnote{Equation~\ref{eq:cross}
corrects eq.~3 of ref.~\cite{mo91}.
Though ref.~\cite{mo91} used eq.~\ref{eq:cross} for calculations,
the $\sigma_{\rm multi-W}^{pp}$ curve in Fig.~1 of ref.~\cite{mo91}
is too large by approximately a factor of 2 due to an incorrect
double-counting of contributions from unlike partons.}
combinations of quarks and antiquarks (but not gluons).
We evaluate all parton distribution functions
at a scale $Q^2 = M_W^2$ and employ the leading-order
parton distributions of Tung and Morfin (fit SL) \cite{TungMorfin}.

In the proton-nucleon center of momentum system (c.m.s.)
where the total energy is $\sqrt{s}$,
we sample the parton distribution functions to generate
the momentum fractions $x_1,x_2$ carried by the quarks
participating in the hard interaction. In the quark-quark c.m.s. the
energy of the hard subprocess is $\sqrt{\hat{s}} = \sqrt{ x_1 x_2 s}.$
For definiteness we assume that multi-W processes produce exactly 30
W bosons. A more detailed scenario should consider the
production of Z bosons
(roughly in the ratio $W^+:W^-:Z  \simeq 1:1:1$), prompt Higgs bosons,
prompt photons and
allow for fluctuations in the total weak boson multiplicity.
In addition, we should, strictly speaking, ensure the conservation
of the quantum numbers carried by the quarks participating in the hard
interaction by allowing more than W bosons in the
final state of the hard subprocess. In the interest of simplicity
we will sacrifice complete consistency and forgo such refinements;
compared to the large number of gauge bosons produced we do not expect
these points to play a significant role in our investigation of
whether or not it is feasible to observe multi-W processes.

We assume that in the quark-quark c.m.s.
the W boson momenta are distributed isotropically with
each W boson carrying an energy $\sqrt{\hat{s}}/30.$ A more detailed
treatment should employ 30-body relativistic phase space and perhaps
impose dynamic assumptions such as limited $p_T \simeq
{\cal O}(m_W)$ of the W bosons
(in analogy with limited $p_T$ in QCD).  However, because of the rapidly
falling parton-parton luminosity, the quark-quark subprocess energy
$\sqrt{\hat{s}}$ tends to lie just above the
multi-W threshold $\sqrt{\hat{s}_0}.$ Consequently,
for multi-W thresholds close
 to
the kinematic limit for the production of 30 W bosons ({\it i.e.,}
$\sqrt{\hat{s}_0} \simeq 2.4~\mbox{TeV}$) there is little extra
energy available to have to
worry about the precise distribution of the W bosons in momentum space.

We employ the Monte Carlo program {\sc jetset} \cite{jetset}
to decay all W bosons and to reproduce
measured W branching fractions and hadronic multiplicities.
At this stage we inhibit the decays of relatively
long-lived secondaries such as
$\pi^\pm,\kappa^\pm,\rho,\eta,\kappa_L,\eta'$ to allow for the
possibility that they may undergo hadronic interactions
with air nuclei in the subsequent air shower. The extensive
air shower simulator described in sect.~\ref{se:showersim}
will determine whether these secondaries decay or interact.

Aside from the decays of the W bosons we must also
consider those parts of the colliding protons which do not
participate in the hard
subprocess --- the so-called spectator fragments.
Since the quarks participating
in the subprocess carry colour,
the hadronization of the spectator
fragments is not, in general, independent of the hadronization
of the subprocess system. As a simplification we ignore this by point
and adopt the following procedure: 1) We treat
the multi-W production subprocess as a color singlet (and hence hadronize
it independently by decaying all 30 W bosons).
2) We replace the spectator fragment originating from the
cosmic proton with a nucleon carrying the same energy.
3) We ignore the spectator fragment originating from the
stationary ``target'' proton. This ansatz, especially steps 2) and 3),
is intended to embody the essential characteristics of the leading
particle effect.
As a final step we boost all W decay products and
the leading nucleon to the Earth rest frame and inject them
into the upper atmosphere to be used as initial conditions
for an extensive air shower.

\subsection{Air shower simulation}

\label{se:showersim}
\hspace*{\parindent}
We use the computer program {\sc showersim}~\cite{showersim}
to simulate both the electromagnetic and hadronic components
of air showers generated by multi-W processes. The program
accounts for multiple hadronic and electromagnetic interactions
in the atmosphere and allows for the decay of
unstable particles. The interested reader is
referred to the {\sc showersim} documentation for a detailed discussion
of the program's physical assumptions.
We employed the program in its default form with few exceptions.
For the electromagnetic components of showers we employed the
{\sc elcas.5} and {\sc tail.2} routines which provide a detailed evolution
of photons  and electrons below 200~GeV;
for the underlying hadronic interaction model we used the
``W00'' option which nominally fits SPS data.

We also used {\sc showersim} to generate samples of
generic air showers induced
by protons and iron nuclei. We stress that though we
employed {\sc showersim} exclusively, we did so only for convenience
since it is not clear that {\it any} available shower model,
especially concerning the hadronic component of extensive air
showers, provides an accurate representation of Nature.
In this respect, our results for multi-W air showers,
generic proton or iron showers may not be accurate on an
absolute scale but should be reliable relative to each
other due to common simulation techniques.

\subsection{Multi-muon detection rates}

\label{se:muon}
\hspace*{\parindent}
For completeness we summarize here the ingredients of
our calculations of multi-muon detection rates. The
interested reader is referred to ref.~\cite{mo91} for
additional details.

We characterize a subsurface detector
by its vertical depth $D$, its geometrical size
(length $L$, width $W$, height $H$) and
a muon threshold energy $E_{\rm thresh}$ which is the
minimum energy required of muon entering the detector
in order that it pass completely through the detector.
For our calculations we idealize DUMAND (with a nine
string array) as a
100~m $\times$ 100~m $\times$ 250~m
volume at an ocean depth of 4.5~km with
$E_{\rm thresh}=100$ GeV. For simplicity we
neglect the effective growth of the array size with
muon energy. Similarly, we idealize MACRO as
77~m$\times$ 12~m $\times$ 9~m volume located
at depth of 3700~hg/cm$^2$ below the surface of
a spherical Earth of density $\rho = 2.6$ g/cm$^3$. For MACRO
we assume $E_{\rm thresh} = 2$ GeV.
Since prompt muons from multi-W phenomena would
typically arrive at
DUMAND or MACRO
with energies $\gwig {\cal O}(10$ TeV),
the small values of $E_{\rm thresh}$ used above are essentially
irrelevant.

For an isotropic differential flux of
cosmic neutrinos, $j_\nu$, the differential flux of detected
events with $k$ muons in coincidence originating from
multi-W phenomena is given by
\begin{equation}
\label{eq:mude}
{dN_{k\mu}\over dA\ dt\ d\Omega} =
\int_{E_{\rm thresh}}^\infty dE\ P_{k\mu} (E,X)\ j_\nu (E),
\end{equation}
where the total column density of matter between the detector
and the upper atmosphere is
\begin{equation}
\label{eq:overburden}
X  = \rho \biggl[
\sqrt{ (R_\oplus - D)^2 \cos^2\theta + 2DR_\oplus - D^2} -
(R_\oplus-D)\cos\theta
\biggr] + X_{\rm atm}.
\end{equation}
The first term in the column density
accounts for rock/water/ice above the detector
and the second term is the appropriate atmospheric slant depth
using the U.S. Standard atmosphere model.

$P_{k\mu}(E,X)$ is the probability that a cosmic neutrino
of energy $E$, initially separated from
the detector by a column density $X,$
gives rise to a multi-W event with $k$ muons detected in
coincidence,
\begin{equation}
\label{eq:pk}
P_{k\mu} (E,X) =
{n_\mu ! \over (n_\mu - k) ! k!}\ N_A
\sigma_{\rm multi-W}^{\nu N}
\int_0^X dX^\prime\ \mbox{e}^{-N_A \sigma_{\rm tot}^{\nu N} (X-X^\prime )}
\
{\tilde p}_\mu^k
(1-{\tilde p}_\mu )^{n_\mu-k} ,
 \end{equation}
where $n_\mu$ is the number of collimated muons produced in a multi-W
event and $N_A=m_p^{-1}$.
We assume $n_\mu = n_W / 9 \simeq 3$,
corresponding to prompt muons from the
decay of 30 W bosons.

$\tilde p_\mu ( E, E_{\rm thresh}, X^\prime )$
is the detection probability
for a typical prompt muon produced in a  multi-W
process induced by a neutrino with energy $E.$ Under the
assumption that the $n_W$ weak gauge
bosons are distributed isotropically in the
subprocess rest frame,
\begin{equation}
\label{eq:tilpefinal}
\tilde p_\mu ( E, E_{\rm thresh}, X^\prime ) =
1 - n_W {E_\mu^\prime (E_{\rm thresh} , X^\prime )\over E} ,
\end{equation}
where $E_\mu^\prime (E_{\rm thresh},X^\prime )$ is the solution of
the muon energy-range relation
\begin{equation}
\label{eq:enrang}
X^\prime = \int_{E_{\rm thresh}}^{E_\mu^\prime}
{dE\over \alpha (E) + \beta (E) E} .
\end{equation}
We neglect the stochastic effects of range straggling
which may become important for muon energies above 10$^5$ GeV
and use parametrizations of $\alpha (E)$ and $\beta(E)$ from
refs. \cite{enloss1,enloss2}, respectively.

The distribution of the inter-muon separation in muon bundles from
neutrino-induced multi-W processes is obtained
as a byproduct of using standard Monte Carlo techniques
to evaluate the integral of eq.~\ref{eq:muonbundlerate}
which gives the number of underground muon bundles.
For each point contributing to the integral of eq.~\ref{eq:muonbundlerate}
one generates a muon bundle configuration, propagates it to the detector
and calculates its contribution to the distribution of pair-wise
separations using the integrand of eq.~\ref{eq:muonbundlerate} as a weight.
Muon bundle configurations are generated according to the assumptions
outlined in sect.~\ref{se:proton}. Namely, for a cosmic neutrino of energy
$E_\nu$ initiating a multi-W process, one samples the parton distribution
functions of the target nucleon to determine
the quark-neutrino c.m.s. energy $\sqrt{\hat{s}} = \sqrt{2 m_p E_\nu x}~
( > \sqrt{\hat{s}_0}).$ Since prompt muons from W boson decay are
distributed
isotropically in the quark-neutrino c.m.s., one generates momentum
vectors for three prompt muons in that frame and then boosts the results
to the Earth rest frame. After including the effects of muon energy loss
and multiple Coulomb scattering in the medium surrounding the detector,
the $2 \le n_\mu \le 3$ muons reaching the detector determine
$n_\mu (n_\mu-1)/2$ pairwise separations which contribute
to the muon separation distribution such as shown in
fig.~\ref{fi:macrodata}.

\newpage
\noindent {\bf \LARGE Figure Captions}

\begin{enumerate}

\item
Universal curves parametrizing multi-W production cross sections
in proton-nucleon ($p N$), proton-electron ($p e^-$),
neutrino-nucleon ($\nu N$) \linebreak and neutrino-electron ($\nu e^-$)
collisions. Curves are for protons and neutrinos with
laboratory energy $E$ colliding with
nucleons and electrons at rest. $E_p^{(pN~{\rm thresh})}
= \hat{s}_0 / ( 2 m_p)$ is the proton threshold energy
for $pN$ multi-W processes. The ($p N$) curve
corrects an error in ref.~\cite{mo91};
the corresponding curve of ref.~\cite{mo91}
is too large by approximately a factor of 2 (see footnote in appendix).
\label{fi:universal}

\item
Contours corresponding to 1 and 10 multi-W events
in one year ($10^7~{\rm s}$) of
operation for the LHC
(${\cal L} = 10^{34}~\mbox{cm}^{-2} ~\mbox{s}^{-1}$)
and the SSC (${\cal L} = 10^{33} ~\mbox{cm}^{-2} ~\mbox{s}^{-1}$).
\label{fi:lhcssc}

\item
Event number contours in $10^7$~s
for proton-induced multi-W air showers
assuming the Constant Mass Composition model for proton flux.
Solid: 100 km$^2$ conventional surface array sensitive to
$E_{\rm shower} \ge 1$ PeV at zenith angles
$\theta \le 60^{\rm o}$. Dashed: Fly's Eye array sensitive to
$E_{\rm shower} \ge 100$~PeV using aperture of ref.~\cite{fly}.
\label{fi:pinitiatedarray}

\item
Differential flux of protons and neutrinos
used in text. The Constant Mass Composition proton flux
is from ref.~\cite{ConstantMassComposition}. The diffuse
neutrino flux from active galactic nuclei (AGN) and
the 2.7 K photoproduced neutrino flux are taken from
ref.~\cite{st91}. Neutrino fluxes shown are
summed over species in the proportion $\nu_\mu:
\bar\nu_\mu:\nu_e:\bar\nu_e = 2:2:1:1$.
\label{fi:pflux}

\item
Differential flux of proton-induced multi-W air showers assuming
the Constant Mass Composition (CMC) model proton flux for
fixed multi-W production parameters.
\label{fi:pinitiatedmultiwflux}

\item
Lateral distributions of electrons $(E_e > 1$ MeV),
muons ($E_\mu > 1$ GeV) and hadrons ($E_{\rm had} > 1$ GeV)
in 5~PeV vertical air showers (lower three curves in each plot) and
30~PeV vertical air showers (upper three curves in each plot)
at atmospheric depth of 800 g/cm$^2.$ Solid curves correspond to
proton-induced multi-W showers assuming $\sqrt{\hat s_0}= 2.4$ TeV
and any value of $\hat\sigma_0.$
Dashed (dot-dashed) curves correspond to generic showers initiated by
proton (iron) primaries. Each curve is an average over 25--100 showers
including variations in the depth of first interaction.
\label{fi:lateraldistsurfacearray}

\item
Average fraction of total shower energy $E_p$
carried by generic component of a proton-induced multi-W air shower.
$E_p^{\rm thresh} = \hat s_0 / ( 2 m_p)$
is the proton threshold energy for multi-W production.
\label{fi:fraction}

\item
Lateral distributions of muons with $E_\mu > 1.5$~TeV
at atmospheric depth of 800 g/cm$^2$ for vertical air showers
initiated by 30 PeV primaries. The multi-W processes assume
$\sqrt{\hat s_0}= 2.4$ TeV. Each curve is an average
over 100--500 showers including variations in the depth of first
interaction.
\label{fi:lateraldisttevmuons}

\item
Longitudinal development curves for 150~PeV hadron-induced
vertical air showers. The depths of first interaction are fixed at
42~g/cm$^2$ for p-induced multi-W showers, 42~g/cm$^2$ for
generic p-initiated showers and
11~g/cm$^2$ for generic Fe-initiated showers.
A multi-W parton-parton threshold of
$\sqrt{\hat s_0} = 5$ TeV is assumed.
A number of curves are shown to illustrate the similarity between
multi-W showers and fluctuations in generic showers.
\label{fi:longdev}

\item
Neutrino interaction length due to combined effects of
generic charged current interactions and multi-W processes.
\label{fi:nuabsorption}

\item
Excluded region (hatched) in $E_\nu-\sigma^{\nu N}_{\rm tot}$
space from combination (solid line) of Fly's Eye limits
with the AGN neutrino flux of Stecker {\it et al.}~\cite{st91}.
Dashed lines indicate limiting
cases of $\sigma_{\rm tot}^{\nu N}$ for
$( \sqrt{\hat s_0} = 8~\mbox{TeV},
   \hat \sigma_0 = .5~\mu\mbox{b} )$
and
$( \sqrt{\hat s_0} = 8~\mbox{TeV},
   \hat \sigma_0 = 81~\mu\mbox{b} )$ which
are consistent with the Fly's Eye limits.
\label{fi:neutrinoinitiatedflyseye1}

\item
Regions of multi-W parameter space excluded by the Fly's Eye.
The region labelled ``AGN $\nu$'' is excluded if one assumes
only the AGN neutrino flux of Stecker {\it et al.}~\cite{st91}.
The region labelled ``2.7 Photoproduced $\nu$'' is
excluded in addition if one includes the neutrino flux
contributions due to the cosmic background radiation
shown in Fig.~\ref{fi:pflux}.
\label{fi:neutrinoinitiatedflyseyeexcluded}

\item
Excluded regions in $E_\nu-\sigma^{\nu N}_{\rm tot}$
space from combination (solid line) of Fly's Eye limits
with the AGN neutrino flux of Stecker {\it et al.}~\cite{st91}
(hatched) and the flux due to the 2.7~K cosmic background
radiation (double hatched) (c.f. Fig.~\ref{fi:pflux}).
Dashed lines indicate limiting cases of $\sigma_{\rm tot}^{\nu N}$
for
$( \sqrt{\hat s_0} = 8~\mbox{TeV},   \hat \sigma_0 = 48~\mbox{nb} )$
and
$( \sqrt{\hat s_0} = 8~\mbox{TeV},   \hat \sigma_0 = 81~\mu\mbox{b} )$
which are consistent with the limits.
\label{fi:neutrinoinitiatedflyseye2}

\item
Event number contours for neutrino-induced multi-W extensive air
showers ($E_{\rm shower} \ge 100$ PeV, zenith
angle $\le 60^{\rm o}$ ) in $10^7$ s for a 100~km$^2$
conventional surface array (vertical depth 1000 g/cm$^2$).
Solid contours includes all showers. Dashed contours include
only showers initiated a minimum of 500 g/cm$^2$ away from array.
The AGN neutrino flux of Stecker {\it et al.}~\cite{st91} is assumed.
\label{fi:neutrinoinitiatedsurfacearray}

\item
Contours for neutrino-induced
contained events in 1~km$^3$ volume of water at
an ocean depth of 4.5~km in $10^7$~s (approximately the
arrangement of the proposed SADCO acoustic detector).
The neutrino flux of Stecker {\it et al.}~\cite{st91} is assumed
(see Fig.~\ref{fi:pflux}).
\label{fi:neutrinoinitiatedcontained}

\item
Contours for neutrino-induced multi-W muon
bundles at zenith angles $\theta > 80^{\rm o}$ in 10$^7$ s
at MACRO and DUMAND assuming the AGN neutrino flux of
Stecker {\it et al.} \cite{st91}.
\label{fi:neutrinoinitiatedmacro}

\item
Contours for neutrino-induced multi-W muon
bundles for all zenith angles in 10$^7$ s
at DUMAND assuming the AGN neutrino flux of
Stecker {\it et al.} \cite{st91} (see Fig.~\ref{fi:pflux}).
\label{fi:dumandallzenith}

\item
Contours for neutrino-induced multi-W muon
bundles for all zenith angles in 10$^7$ s
at MACRO assuming the AGN neutrino flux of
Stecker {\it et al.} \cite{st91}.
\label{fi:neutrinoinitiatedallzenith}

\item
MACRO pairwise muon separation data \cite{bun92} compared with
expectations for neutrino-induced multi-W phenomena
for $(\sqrt{\hat s_0} = 2.4$ TeV , $\hat\sigma_0 = 10~\mu$b)
assuming the AGN neutrino flux of Stecker {\it et al.} \cite{st91}.
MACRO data corresponds to two supermodules operating for 2334.3 hours
sensitive to bundles with zenith angle $\theta < 60^{\rm o}.$
\label{fi:macrodata}

\end{enumerate}
\newpage
\oddsidemargin=1cm
\begin{center}
\begin{description}
\item[Table 1]
Average particle multiplicity at atmospheric depth of
$800~\mbox{g/cm}^2$ for vertical extensive air showers generated
by 5~PeV (and 30~PeV) primaries. Multi-W showers are proton-induced.
Showers labelled p and Fe contain only generic interactions.
\end{description}
\end{center}

\begin{center}
\begin{tabular}
  { | c |  c | c | c | c | } \hline
               &                 &                 &                &   \\
Shower    & $\langle N_e         \rangle$     &
                 $\langle N_\mu       \rangle$     &
                 $\langle N_{\rm had} \rangle$     &
                 $\langle N_\mu       \rangle$      \\
Type               & $E_e         > 1  ~\mbox{MeV}  $ &
                 $E_\mu       > 1  ~\mbox{GeV}  $ &
                 $E_{\rm had} > 1  ~\mbox{GeV}  $ &
                 $E_\mu       > 1.5~\mbox{TeV}  $ \\
               &                 &                 &                &   \\
\hline
               &                 &                 &                &   \\
{\rm multi-W}  & $1.1 \times 10^6$ & $5.9 \times 10^4$ & $2.2\times 10^3$ &
18\\
               & ($1.4 \times 10^7$) &
($1.9 \times 10^5$) & ($1.4\times 10^5$) & (40)\\
               &                   &               &                &    \\
p              & $2.4 \times 10^6$ & $4.5 \times 10^4$ & $2.9\times 10^3$ & 5
 \\
               & ($1.8 \times 10^7)$ & ($1.5 \times 10^5$) & ($1.5\times
10^5$)
 & (20)\\
               &                 &                 &                &    \\
Fe             & $1.3 \times 10^6$ & $8.7 \times 10^4$ & $3.3\times 10^3$ &
14\\
               & ($1.1 \times 10^7$) & ($3.3 \times 10^5$) & ($1.4\times 10^5$)
 & (60)\\
               &                 &                 &                &    \\
\hline
\end{tabular}
\end{center}

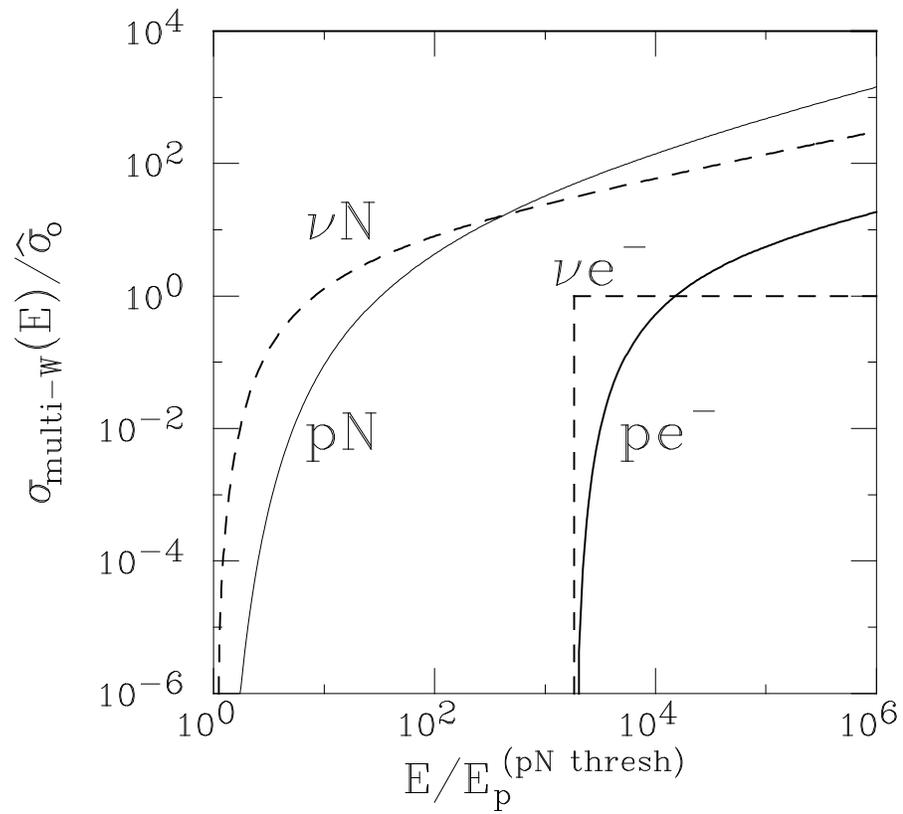
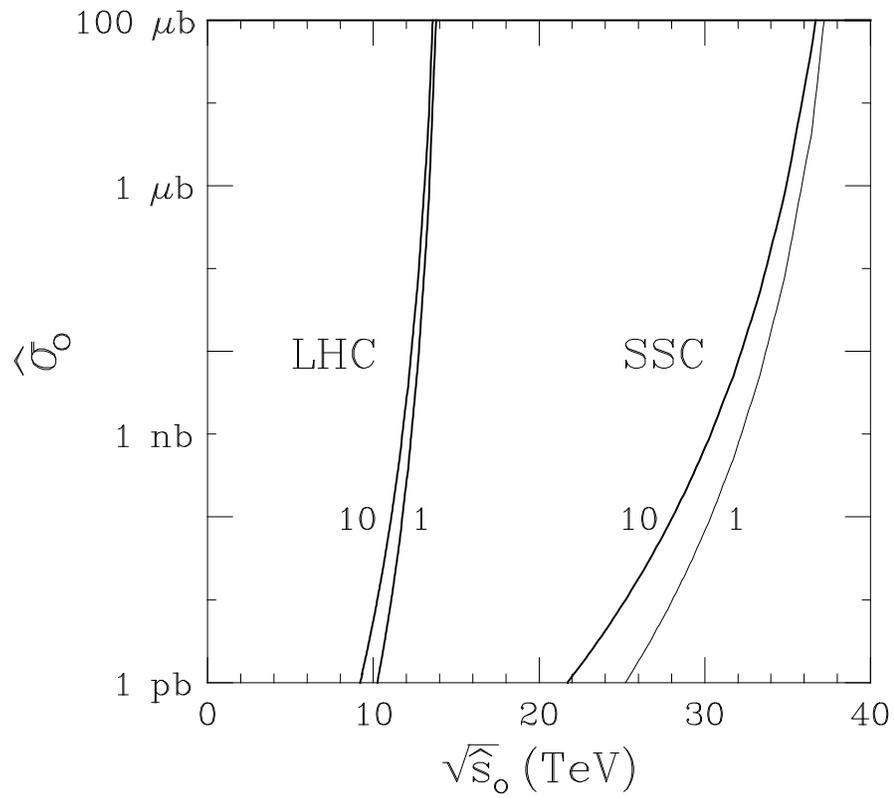
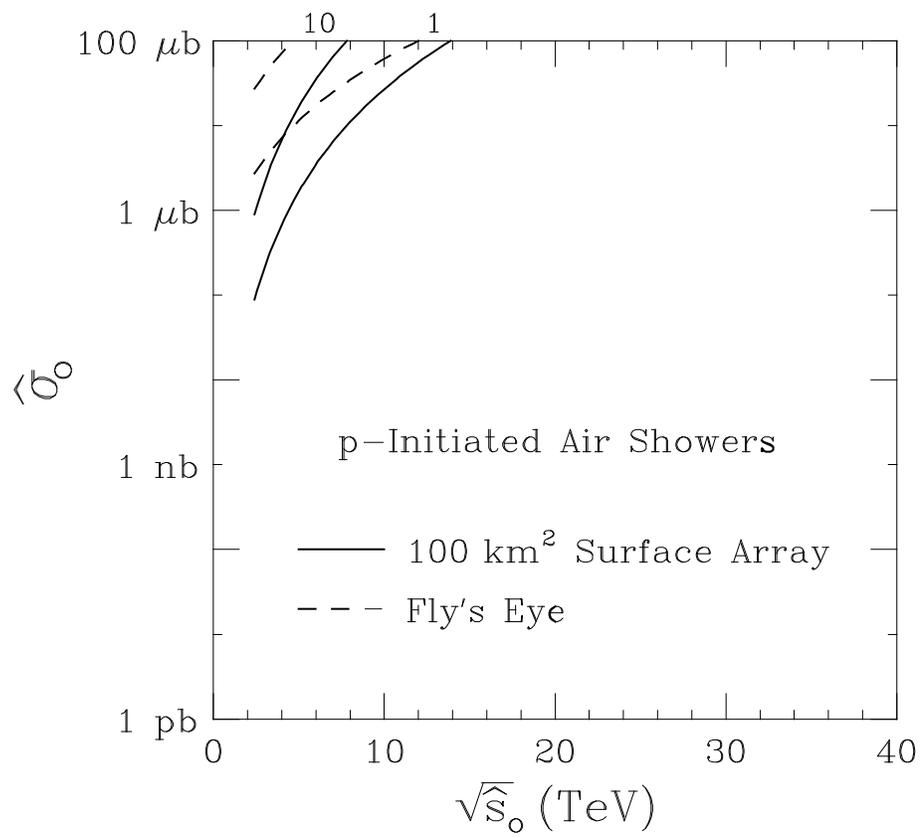
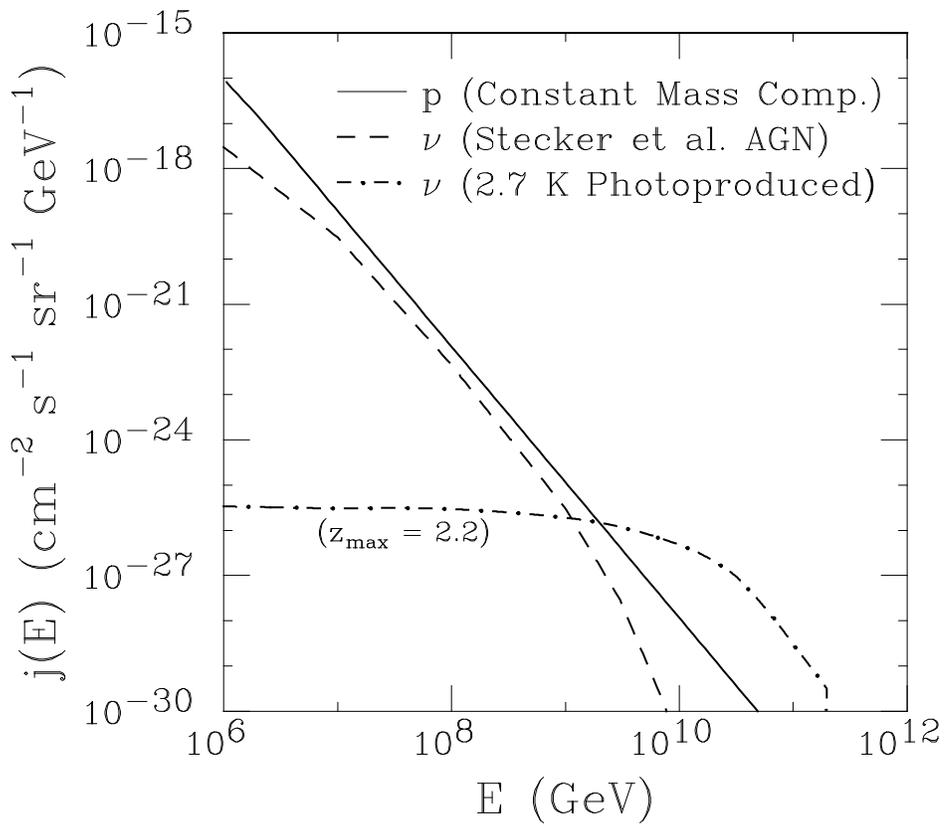
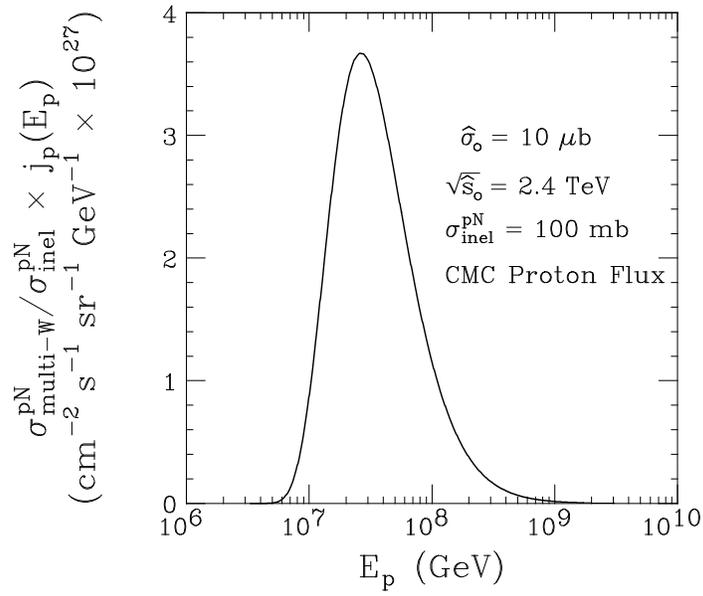
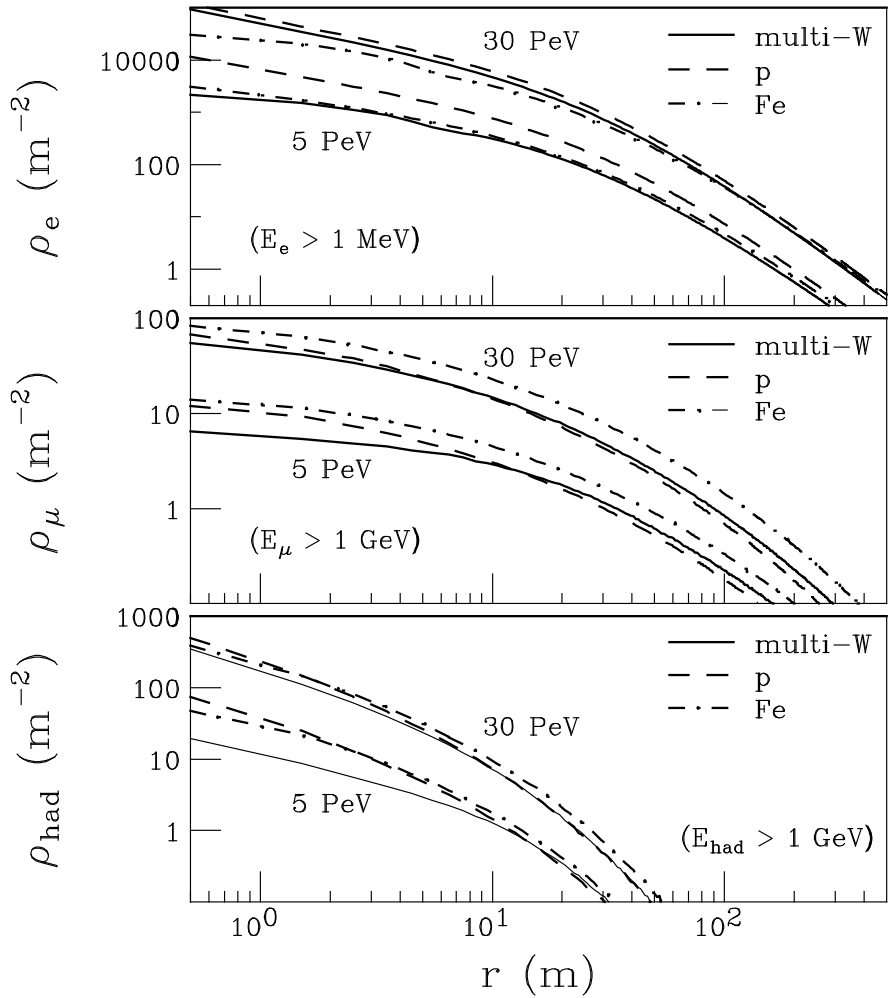
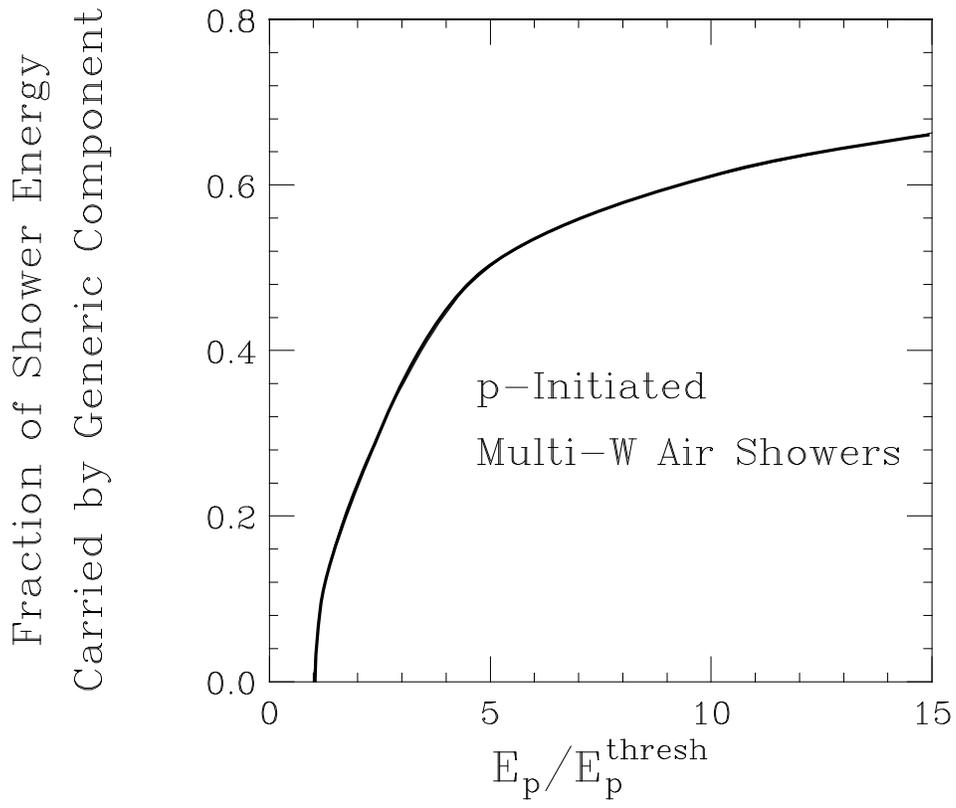
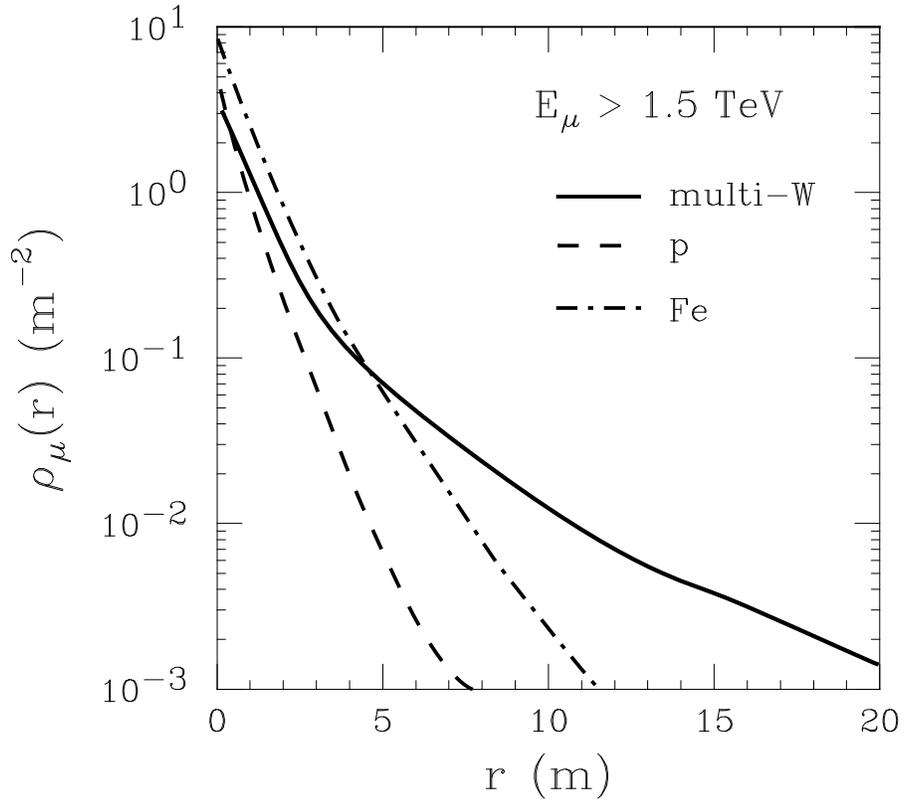
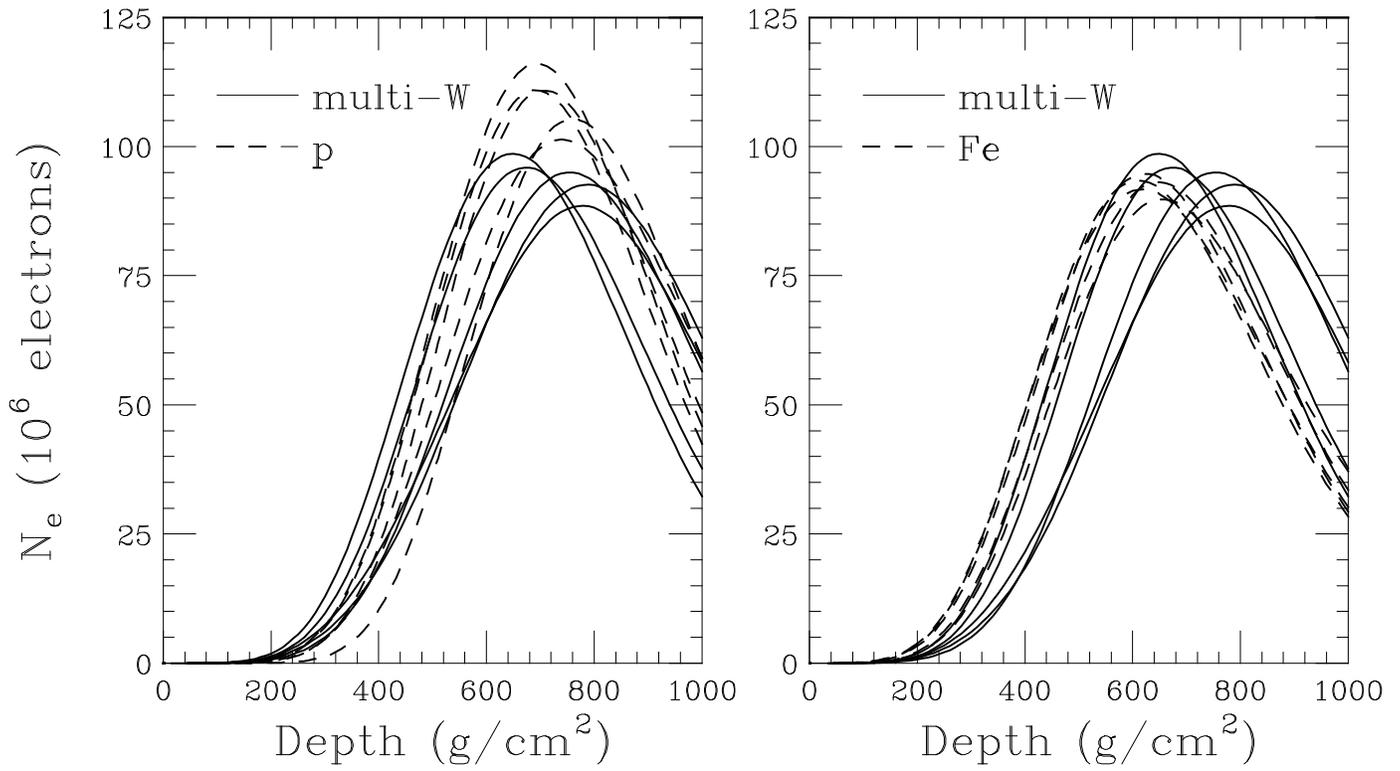
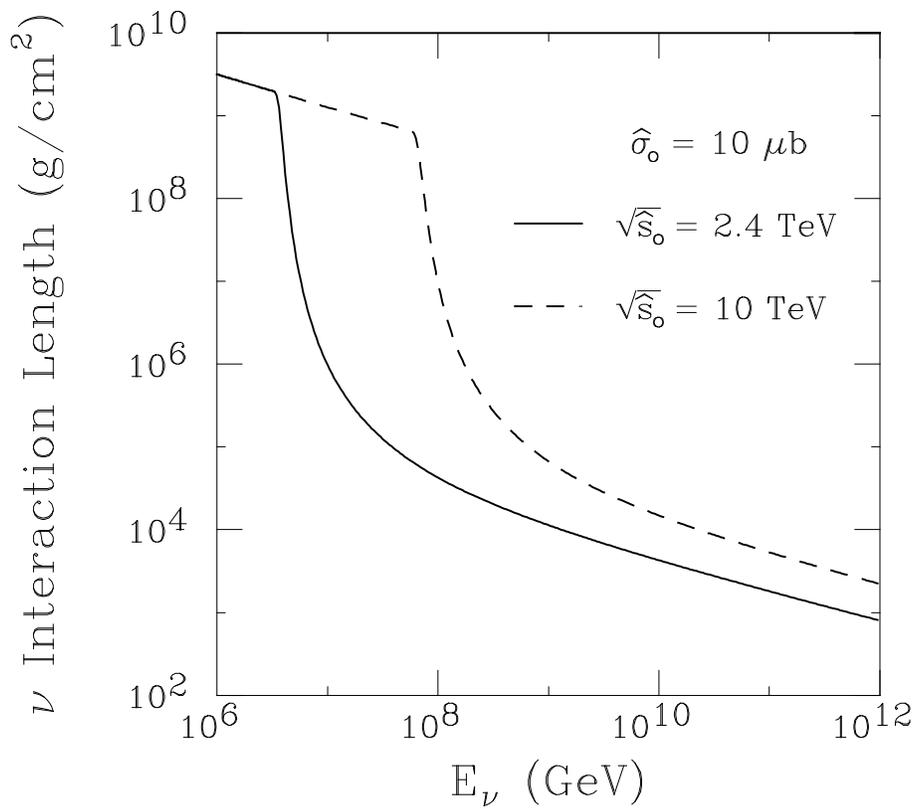
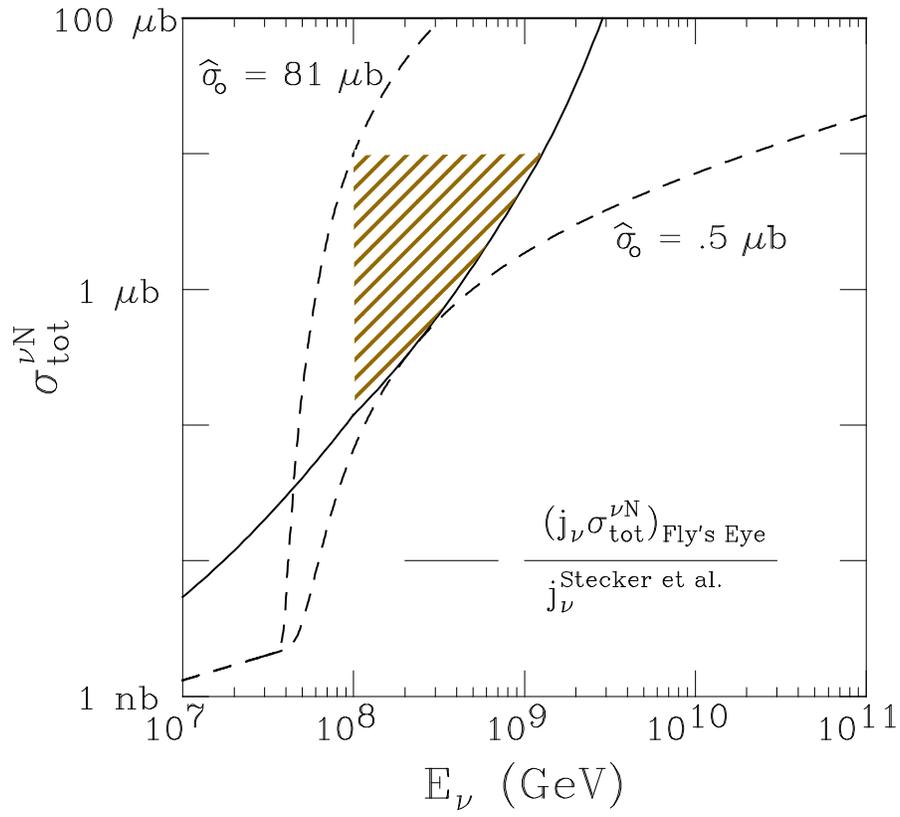
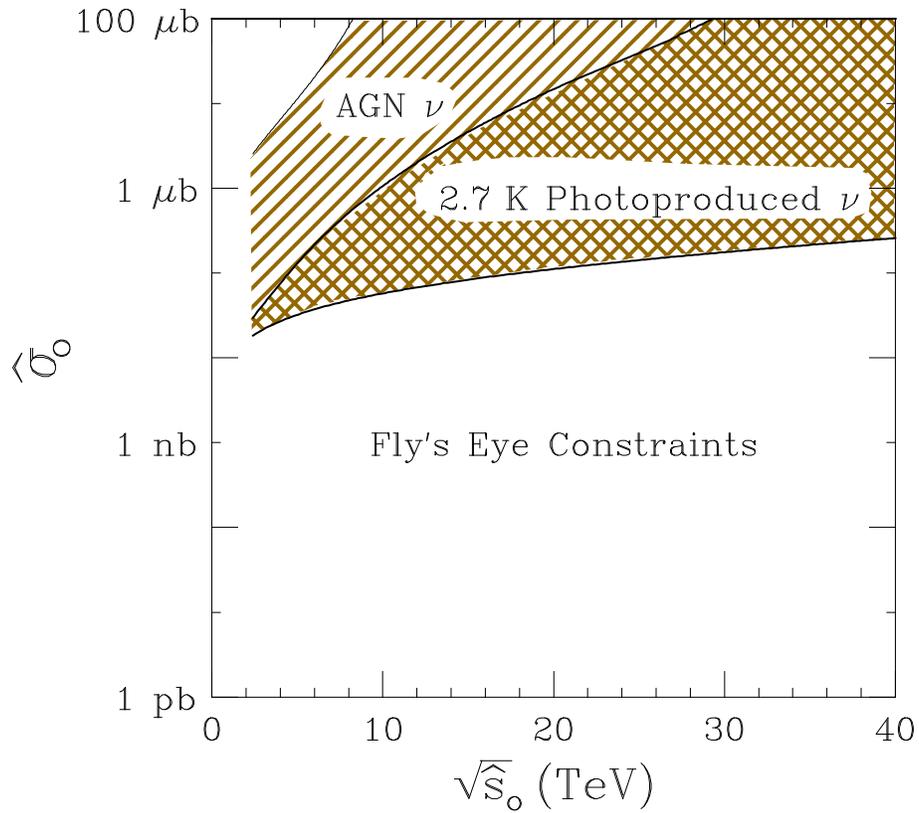
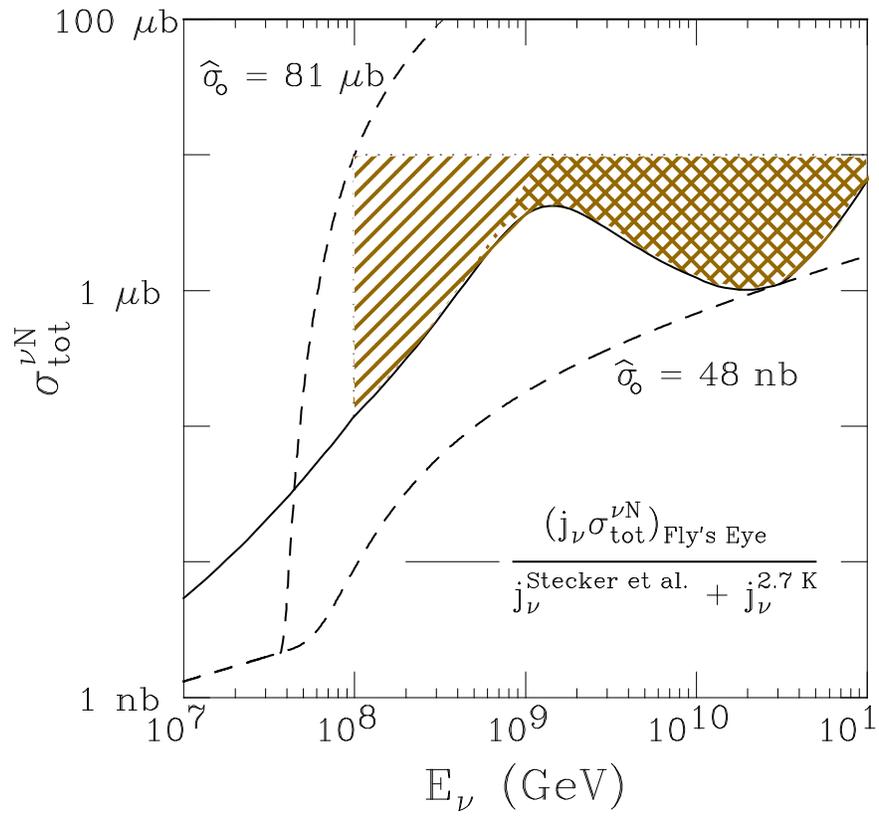
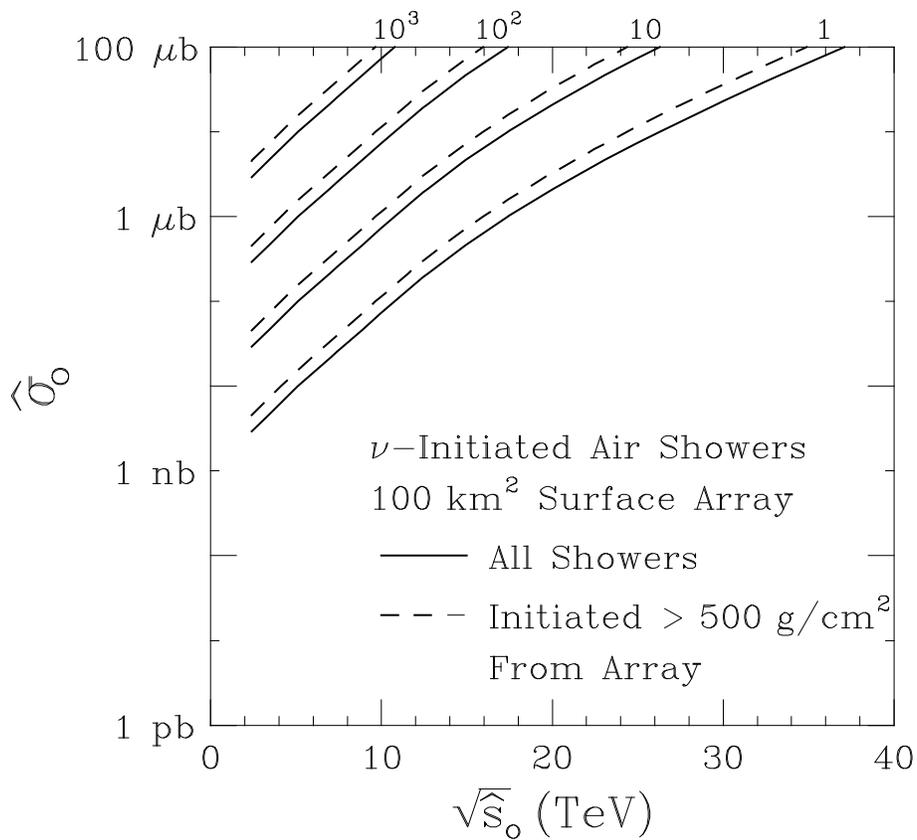
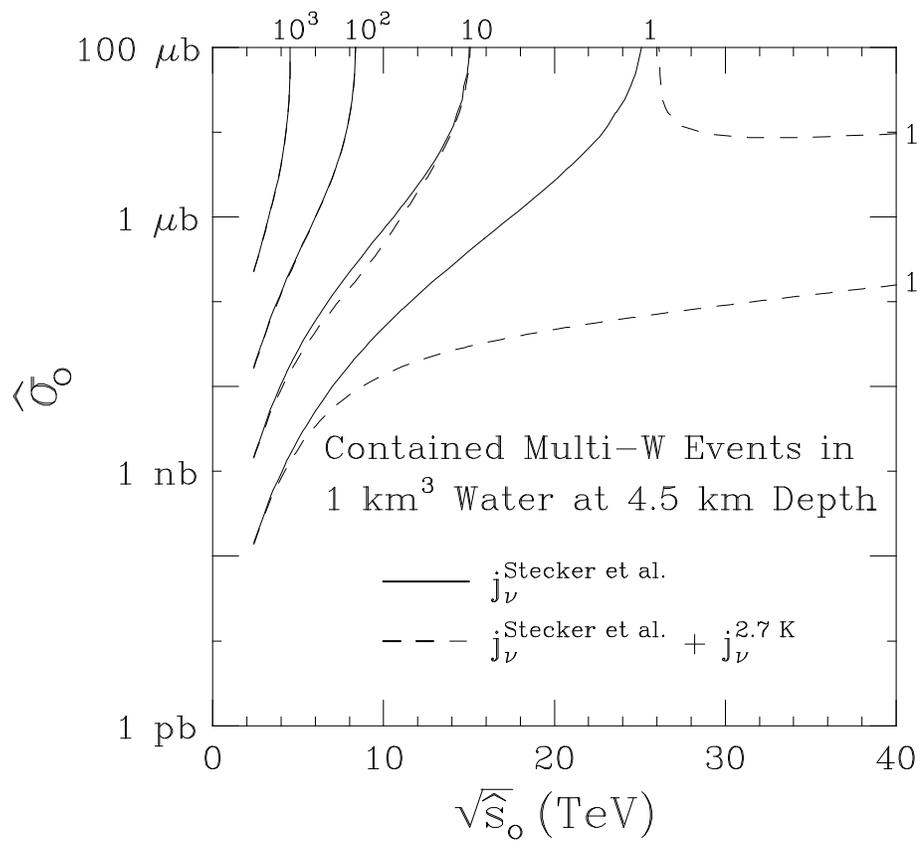
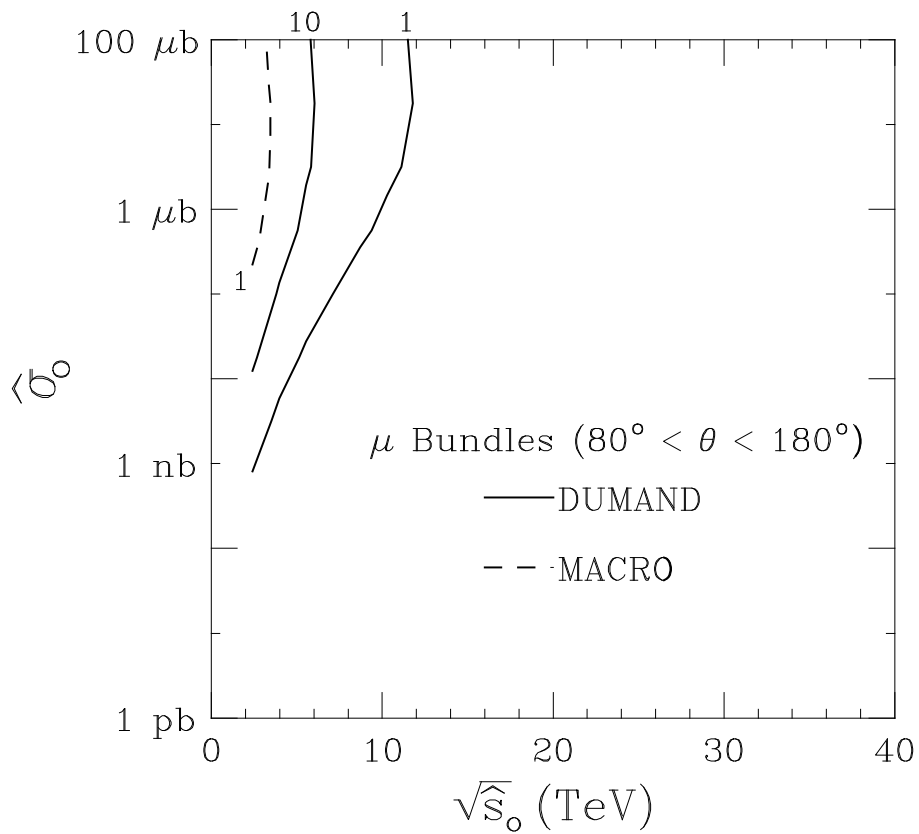
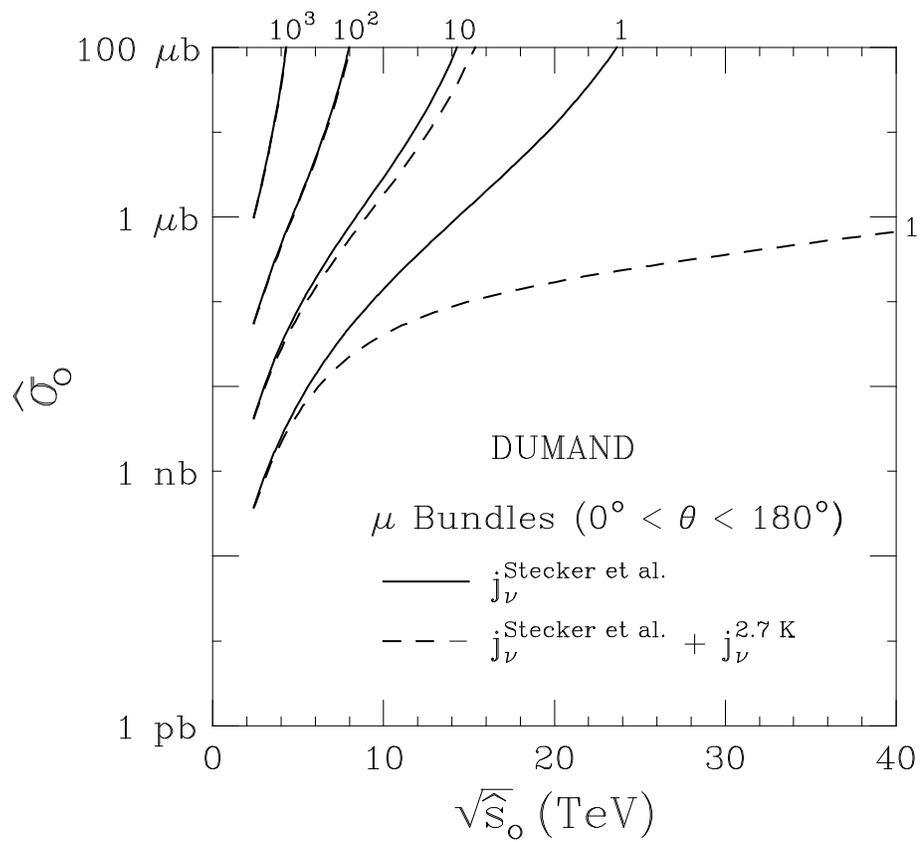
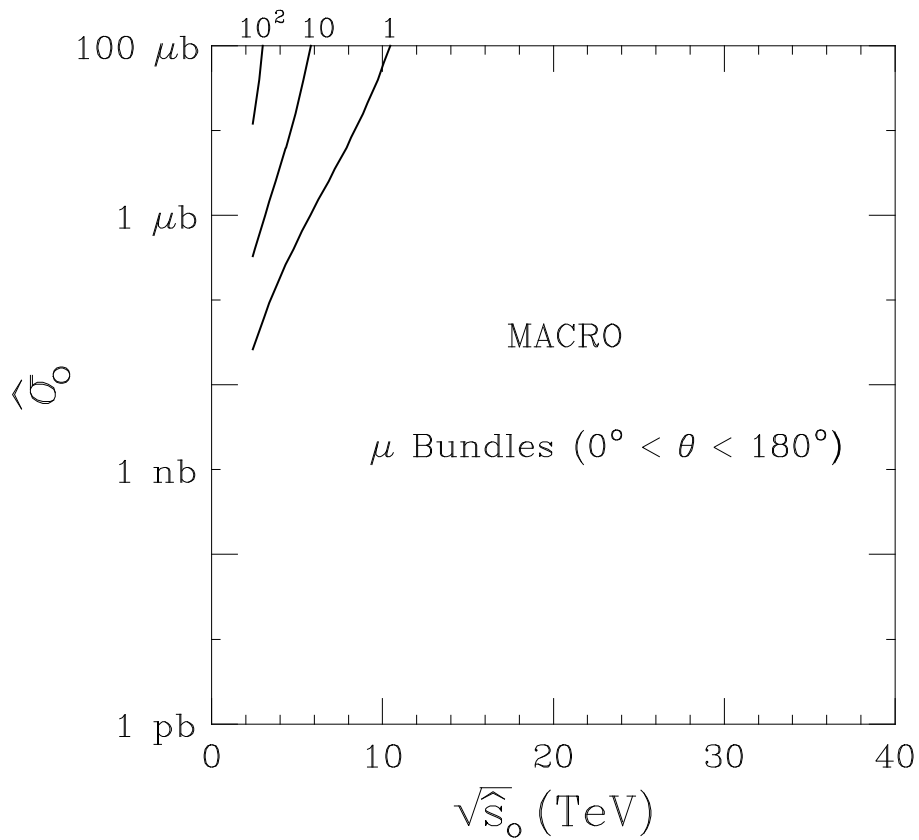
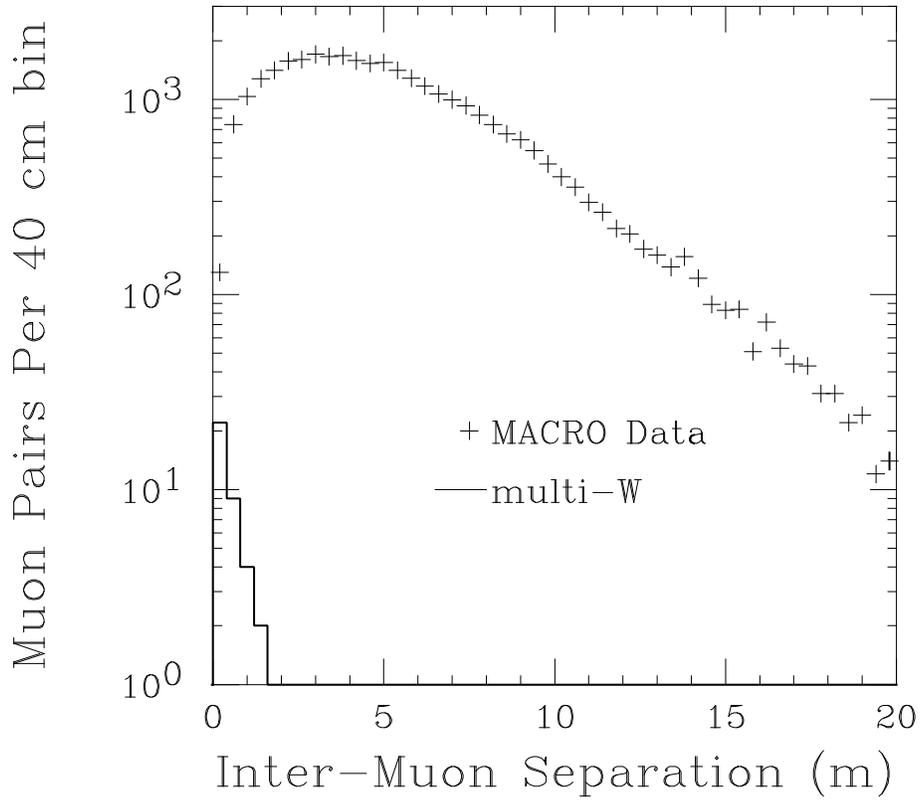
\end{document}